\documentclass[fleqn,usenatbib]{mnras}

\newcommand{\PVdblt}{{\rm P}\kern 0.1em{\sc v}~$\lambda\lambda 1117, 1128$}
\newcommand{\CaIIdblt}{{\rm Ca}\kern 0.1em{\sc ii}~$\lambda\lambda 3934, 3969$}
\newcommand{\AlIIIdblt}{{\rm Al}\kern 0.1em{\sc iv}~$\lambda\lambda 1855, 1863$}
\newcommand{\CIVdblt}{{\rm C}\kern 0.1em{\sc iv}~$\lambda\lambda 1548, 1550$}
\newcommand{\MgIIdblt}{{\rm Mg}\kern 0.1em{\sc ii}~$\lambda\lambda 2796, 2803$}
\newcommand{\NVdblt}{{\rm N}\kern 0.1em{\sc v}~$\lambda\lambda 1238, 1242$}  
\newcommand{\SVIdblt}{{\rm S}\kern 0.1em{\sc vi}~$\lambda\lambda 933, 944$} 
\newcommand{\OVIdblt}{{\rm O}\kern 0.1em{\sc vi}~$\lambda\lambda 1031, 1037$} 
\newcommand{\SiIIdblt}{{\rm Si}\kern 0.1em{\sc ii}~$\lambda\lambda 1190, 1193$} 
\newcommand{\SiIVdblt}{{\rm Si}\kern 0.1em{\sc iv}~$\lambda\lambda 1393, 1402$} 
\newcommand{\PV}{\hbox{{\rm P}\kern 0.1em{\sc v}}}
\newcommand{\AlI}{\hbox{{\rm Al}\kern 0.1em{\sc i}}}
\newcommand{\AlII}{\hbox{{\rm Al}\kern 0.1em{\sc ii}}}
\newcommand{\AlIII}{{\hbox{\rm Al}\kern 0.1em{\sc iii}}}
\newcommand{\CaII}{\hbox{{\rm Ca}\kern 0.1em{\sc ii}}}
\newcommand{\CII}{\hbox{{\rm C}\kern 0.1em{\sc ii}}}
\newcommand{\CIIe}{\hbox{{\rm C$^{\ast}$}\kern 0.1em{\sc ii}}}
\newcommand{\CIII}{\hbox{{\rm C}\kern 0.1em{\sc iii}}}
\newcommand{\CIV}{\hbox{{\rm C}\kern 0.1em{\sc iv}}}
\newcommand{\CV}{\hbox{{\rm C}\kern 0.1em{\sc v}}}
\newcommand{\HI}{\hbox{{\rm H}\kern 0.1em{\sc i}}}
\newcommand{\HII}{\hbox{{\rm H}\kern 0.1em{\sc ii}}}
\newcommand{\Lya}{\hbox{{\rm Ly}\kern 0.1em$\alpha$}}
\newcommand{\Lyb}{\hbox{{\rm Ly}\kern 0.1em$\beta$}}
\newcommand{\Lyg}{\hbox{{\rm Ly}\kern 0.1em$\gamma$}}
\newcommand{\Lyd}{\hbox{{\rm Ly}\kern 0.1em$\delta$}}
\newcommand{\Lye}{\hbox{{\rm Ly}\kern 0.1em$\epsilon$}}
\newcommand{\Lyphi}{\hbox{{\rm Ly}\kern 0.1em$\Phi$}}
\newcommand{\Lyfive}{\hbox{{\rm Ly}\kern 0.1em$5$}}
\newcommand{\Lysix}{\hbox{{\rm Ly}\kern 0.1em$6$}}
\newcommand{\Lyseven}{\hbox{{\rm Ly}\kern 0.1em$7$}}
\newcommand{\Lyeight}{\hbox{{\rm Ly}\kern 0.1em$8$}}
\newcommand{\Lynine}{\hbox{{\rm Ly}\kern 0.1em$9$}}
\newcommand{\Lyten}{\hbox{{\rm Ly}\kern 0.1em$10$}}
\newcommand{\Lyeleven}{\hbox{{\rm Ly}\kern 0.1em$11$}}
\newcommand{\HeI}{\hbox{{\rm He}\kern 0.1em{\sc i}}}
\newcommand{\HeII}{\hbox{{\rm He}\kern 0.1em{\sc ii}}}
\newcommand{\FeI}{\hbox{{\rm Fe}\kern 0.1em{\sc i}}}
\newcommand{\FeII}{\hbox{{\rm Fe}\kern 0.1em{\sc ii}}}
\newcommand{\FeIII}{\hbox{{\rm Fe}\kern 0.1em{\sc iii}}}
\newcommand{\MnII}{\hbox{{\rm Mn}\kern 0.1em{\sc ii}}}
\newcommand{\MgI}{\hbox{{\rm Mg}\kern 0.1em{\sc i}}}
\newcommand{\MgIb}{\hbox{{\rm Mg}\kern 0.1em{\sc i}}\kern 0.05em{\rm b}}
\newcommand{\MgII}{\hbox{{\rm Mg}\kern 0.1em{\sc ii}}}
\newcommand{\MgIII}{\hbox{{\rm Mg}\kern 0.1em{\sc iii}}}
\newcommand{\NI}{\hbox{{\rm N}\kern 0.1em{\sc i}}}
\newcommand{\NII}{\hbox{{\rm N}\kern 0.1em{\sc ii}}}
\newcommand{\NIII}{\hbox{{\rm N}\kern 0.1em{\sc iii}}}
\newcommand{\NV}{\hbox{{\rm N}\kern 0.1em{\sc v}}}
\newcommand{\OVI}{\hbox{{\rm O}\kern 0.1em{\sc vi}}}
\newcommand{\OI}{\hbox{{\rm O}\kern 0.1em{\sc i}}}
\newcommand{\OII}{\hbox{[{\rm O}\kern 0.1em{\sc ii}]}}
\newcommand{\OIII}{\hbox{[{\rm O}\kern 0.1em{\sc iii}]}}
\newcommand{\OIV}{\hbox{{\rm O}\kern 0.1em{\sc iv}]}}
\newcommand{\SI}{{\rm S}\kern 0.1em{\sc i}}
\newcommand{\SIV}{{\rm S}\kern 0.1em{\sc iv}}
\newcommand{\SVI}{{\rm S}\kern 0.1em{\sc vi}}
\newcommand{\SiI}{\hbox{{\rm Si}\kern 0.1em{\sc i}}}
\newcommand{\SiII}{\hbox{{\rm Si}\kern 0.1em{\sc ii}}}
\newcommand{\SiIII}{\hbox{{\rm Si}\kern 0.1em{\sc iii}}}
\newcommand{\SiIV}{\hbox{{\rm Si}\kern 0.1em{\sc iv}}}
\newcommand{\SII}{\hbox{{\rm S}\kern 0.1em{\sc ii}}}
\newcommand{\SIII}{\hbox{{\rm S}\kern 0.1em{\sc iii}}}
\newcommand{\NaI}{\hbox{{\rm Na}\kern 0.1em{\sc i}}}
\newcommand{\NaID}{\hbox{{\rm Na}\kern 0.1em{\sc i}}\kern 0.05em{\rm D}}
\newcommand{\TiII}{\hbox{{\rm Ti}\kern 0.1em{\sc ii}}}
\newcommand{\kms}{\hbox{~km~s$^{-1}$}}

\newcommand{\Rv}{\hbox{$R_{\rm vir}$}}
\newcommand{\mfghs}{the Multiphase Galaxy Halos Survey}
\usepackage{newtxtext,newtxmath}

\usepackage[T1]{fontenc}

\DeclareRobustCommand{\VAN}[3]{#2}
\let\VANthebibliography\thebibliography
\def\thebibliography{\DeclareRobustCommand{\VAN}[3]{##3}\VANthebibliography}

\usepackage{graphicx}	
\usepackage{amsmath}	

\title[Chemodynamical structure of the CGM]{The chemodynamical signature of coherent metal-poor inflow and enriched recycled accretion in the cool circumgalactic medium}

\author[G. G. Kacprzak et al.]{Glenn G. Kacprzak$^{1}$\thanks{E-mail: gkacprzak@swin.edu.au},
Jerrard Doran$^{1}$,
Sameer$^{2}$, 
James Farrington$^{1}$,
Jane C. Charlton$^{3}$,
Nikole M. Nielsen$^{2}$, 
\newauthor Kaustubh R. Gupta$^{1}$, Christopher W. Churchill$^{4}$, Tania M. Barone$^{1,5}$, Antonia Fern\'{a}ndez-Figueroa$^{1}$
\\
$^{1}$Centre for Astrophysics and Supercomputing, Swinburne University of Technology, Hawthorn, Victoria 3122, Australia\\
$^{2}$Homer L. Dodge Department of Physics and Astronomy, The University of Oklahoma, 440 W. Brooks St., Norman, OK 73019, USA\\
$^{3}$Department of Astronomy and Astrophysics, The Pennsylvania State University, State College, PA 16801, USA\\
$^{4}$Department of Astronomy, New Mexico State University, Las Cruces, NM 88003, USA\\
$^{5}$Center for Astrophysics, Harvard \& Smithsonian, Cambridge, MA 02138, USA\\
}

\date{Accepted 2026 July 14. Received 2026 June 25; in original form 2026 March 26.}

\pubyear{\the\year{}}

\begin{document}
\label{firstpage}
\pagerange{\pageref{firstpage}--\pageref{lastpage}}
\maketitle

\begin{abstract}
The azimuthal and kinematic structure of the CGM is often interpreted as planar accretion and bipolar outflows, yet direct metallicity evidence for this picture remains ambiguous. We combine cloud-by-cloud ionisation modelling with galaxy rotation kinematics for 21 galaxies from {\mfghs} to investigate how metallicity depends on azimuthal angle and angular momentum. We find that low-ionisation clouds kinematically consistent with disk rotation have $\approx0.5$~dex lower metallicity near the projected major axis ($\Phi<30^\circ$) than at larger azimuthal angles. Major-axis clouds also exhibit higher $N({\HI})$, higher density, and reduced non-thermal line broadening compared to clouds at larger azimuthal angles. In contrast, the higher-ionisation phase shows no significant metallicity dependence on azimuthal angle and has lower column densities, lower densities, higher temperatures, and broader line widths than the co-rotating major-axis low-ionisation clouds. These combined metallicity--kinematic--ionisation signatures are consistent with dynamically cold, metal-poor inflow along the disk plane and enriched, more turbulent gas at larger azimuthal angles that likely traces angular-momentum-supported recycled accretion, embedded within a dynamically complex warmer phase. These results show that metallicity and angular momentum are jointly imprinted by the baryon cycle and are both required to uncover the physical origins of CGM gas.
\end{abstract}





\section{Introduction}

The circumgalactic medium (CGM) is thought to regulate galaxy growth through a combination of accretion and feedback-driven outflows \citep{tumlinson17,fauchergiguere23}. Hydrodynamical simulations predict a geometrically structured halo in which metal-poor inflows preferentially align with the disk plane while enriched winds emerge along the minor axis \citep{peroux20,defilippis21,pillepich21,hafen22,trapp22,yang24}. Observational surveys have established an azimuthal bimodality in absorber incidence and equivalent width for ions such as {\MgII} \citep{bordoloi11,bouche12,kacprzak12,lan14,martin19,zabl19,lundgren21}, {\OVI} \citep{kacprzak15,beckett21,dutta26}, and {\HI} \citep{beckett21,dutta26}. However, direct metallicity evidence for distinct inflow and outflow channels has remained more ambiguous.

Simulations predict that CGM metallicity should be enhanced along the minor axis compared with the major axis, with angular metallicity gradients persisting across a broad range of galaxy masses at $z<1$ \citep{peroux20,vandevoort21}. However, observational evidence for such gradients is mixed. The MEGAFLOW survey measured dust depletion using [Zn/Fe] and found that the CGM along the minor axis is more metal enriched by $\sim$1~dex than the gas along the major axis, assuming dust depletion traces metallicity \citep{wendt21}. In contrast, the Multiphase Galaxy Halos Survey (MGHS) concluded that while the cool CGM metallicity distribution may be bimodal, it shows no dependence on azimuthal angle \citep{pointon19}. Even when accounting for differences between galaxy interstellar medium metallicities and CGM metallicities, no clear azimuthal dependence was found \citep{peroux16,prochaska17, kacprzak19metals}. However, since these studies typically assign a single metallicity to multi-component absorption systems, sightline-averaged analyses inherently mix distinct gas phases and flow histories, and possibly suppress any angular metallicity structure \citep{churchill15,liang18,peeples19,marra21}.

In an effort to disentangle metal mixing in individual absorption systems, a forward-modelling cloud-by-cloud Multiphase Bayesian Ionization Modelling method was developed \citep{sameer21,sameer22}.  Using this approach, \citet{sameer24} examined 47 galaxies from the MGHS and showed that the number of clouds per sightline is enhanced near both the major and minor axes, yet individual cloud metallicities show no dependence on azimuthal angle or other galaxy properties. They concluded that both accretion and outflows contribute to CGM cloud populations, and that individual sightlines probe gas of mixed origin at all azimuthal angles.

Complementary insight into CGM gas origins has come from kinematic studies comparing CGM absorption velocities with galaxy rotation curves. A substantial fraction of {\MgII} absorption exhibits velocities consistent with the direction of galaxy rotation and is broadly consistent with gas accretion models \citep{steidel02,kacprzak10,kacprzak11kin,bouche13,bouche16,diamond-stanic16,ho17,rahmani18,martin19,lopez20}. More recent surveys demonstrate that kinematic behaviour depends on gas phase and galactocentric radius: low-ionisation gas in the inner halo frequently co-rotates with the disk at sub-centrifugal velocities, consistent with spiralling inflow, whereas high-ionisation gas becomes increasingly kinematically decoupled at larger radii \citep{nateghi24GFII,kacprzak19kine,kacprzak25,ho26}.
A similar phase dependence is seen in recent IllustrisTNG simulations of Milky Way analogs, which show that cool gas is more strongly aligned with the disk angular momentum axis than warm gas, and that the co-rotation fraction declines systematically from low to high ions \citep{messere26}.

These observational trends align with simulations predicting that cool inflowing gas retains significant angular momentum and co-rotates with galactic disks, while hotter gas associated with outflows and recycled material exhibits more complex, partially decoupled kinematics \citep{stewart11,danovich15,hafen22}. Simulations further indicate that a substantial fraction of accreting gas is recycled through galactic fountains, in which enriched outflows cool and reaccrete as co-rotating inflows \citep{angles17,muratov17,marinacci19,stern24}. Because recycled gas originates in enriched outflows, co-rotating inflow near the disk is expected to span a wide metallicity range, with metal-poor gas tracing relatively pristine filamentary accretion and metal-rich gas tracing recycled fountains \citep{hafen19,hafen22,trapp22,peroux20,weng24}. This suggests that combining metallicity with kinematics provides a powerful means of distinguishing between pristine inflow, recycled accretion and outflowing gas.

Motivated by these predictions, \citet{nateghi24GFII} demonstrated that combining kinematics and metallicity can provide a sensitive probe of CGM gas flows. For a few galaxies, they showed that low-ionisation CGM gas with high co-rotation fractions tends to be metal poor, consistent with cold accretion along the major axis, whereas more metal-rich gas exhibits weaker kinematic coupling and likely traces recycled accretion or outflows. These results imply that any azimuthal metallicity signature may only emerge once the CGM is separated by both ionisation state and kinematic~behaviour.

In this Letter, we revisit the azimuthal distribution of CGM metallicity using the MGHS cloud catalogue of \citet{sameer24}, and classify individual absorption clouds by ionisation state and by whether their velocities are consistent with host galaxy rotation. Section~2 describes the data and classification, Section~3 presents the results, and Section~4 discusses implications for inflows, outflows and the baryon cycle. Throughout, we adopt a flat $\Lambda$CDM cosmology with $H_{0} = 70~{\rm km~s^{-1}~Mpc^{-1}}$, $\Omega_{\rm m} = 0.3$, and $\Omega_{\Lambda} = 0.7$.

\section{The Sample}
The MGHS comprises 47 galaxy--quasar pairs at redshifts $0.08 \lesssim z \lesssim 0.7$. Spectra from HST/COS \citep{green12}, Keck/HIRES \citep{vogt-hires} and VLT/UVES \citep{dekker-uves} show absorption in a suite of low- and high-ionisation transitions. \citet{sameer24} performed detailed ionisation modelling for each absorption component using \textsc{cloudy} and with nested sampling Bayesian inference techniques, identifying three populations of clouds: cool clouds in photo-ionisation equilibrium (PIE), warm--hot clouds undergoing rapid cooling and requiring time-dependent photo-ionisation (TDP), and hot collisionally ionised gas. The TDP clouds were further subclassified into two types. TDP-high clouds show absorption only in high-ionisation species, at least {\OVI} and possibly {\NV} and/or {\CIV}, with no detected low- or intermediate-ionisation absorption. TDP-low clouds show absorption in intermediate- and high-ionisation species, and in some cases also in low-ionisation states. For each cloud, \citet{sameer24} inferred the total hydrogen column density, density, temperature and metallicity.

We cross-matched this sample with the galaxy rotation kinematics from \citet{nateghi24GFI,nateghi24GFII}, yielding 21 galaxies with rotation curves in common with \citet{sameer24} that have $i> 30$ degrees, where reliable azimuthal angles can be measured. These galaxies span redshifts $0.08 \leq z \leq 0.5$ ($\langle z \rangle = 0.25 \pm 0.11$), a stellar mass range of $8.92 \leq \log(M_\star/M_{\odot}) \leq 10.88$ ($\langle \log(M_\star/M_{\odot}) \rangle = 10.15 \pm 0.59$), an impact parameter range of $21 \leq D \leq 174$~kpc ($\langle D \rangle = 72 \pm 36$~kpc) and a virial radii normalised impact parameter range of $0.16 \leq D/\Rv \leq 1.13$ ($\langle D/\Rv \rangle = 0.53 \pm 0.29$). These 21 galaxies are associated with a total of 63 PIE, 14 TDP-low and 27 TDP-high clouds.\footnote{TDP-high clouds have only metallicity upper limits and are not used in this analysis.}

\section{Results}
We examine the azimuthal distribution of CGM cloud metallicities by combining the cloud-by-cloud ionisation modelling of \citet{sameer24} with the rotation-curve kinematic classifications of \citet{nateghi24GFI,nateghi24GFII}. Clouds are divided by ionisation phase (PIE versus TDP-low) and by whether their velocities are consistent with galaxy rotation. Rotation-consistent implies that the cloud's line-of-sight velocity is consistent with the direction of host galaxy rotation towards the quasar sightline. Each galaxy may have multiple associated clouds, with some being kinematically consistent and some being kinematically inconsistent. For quantities containing upper limits, such as metallicity and hydrogen density, we estimate sample means using Kaplan–Meier survival analysis \citep{kaplan58}, treating limits as left-censored measurements. Uncertainties are derived via bootstrap resampling (5000 realisations), where we report the standard error of the mean estimated as the standard deviation of the bootstrap mean distribution. Differences between subsamples are evaluated using the same bootstrap procedure.

\begin{figure*}
    \centering
    \includegraphics[width=0.495\textwidth]{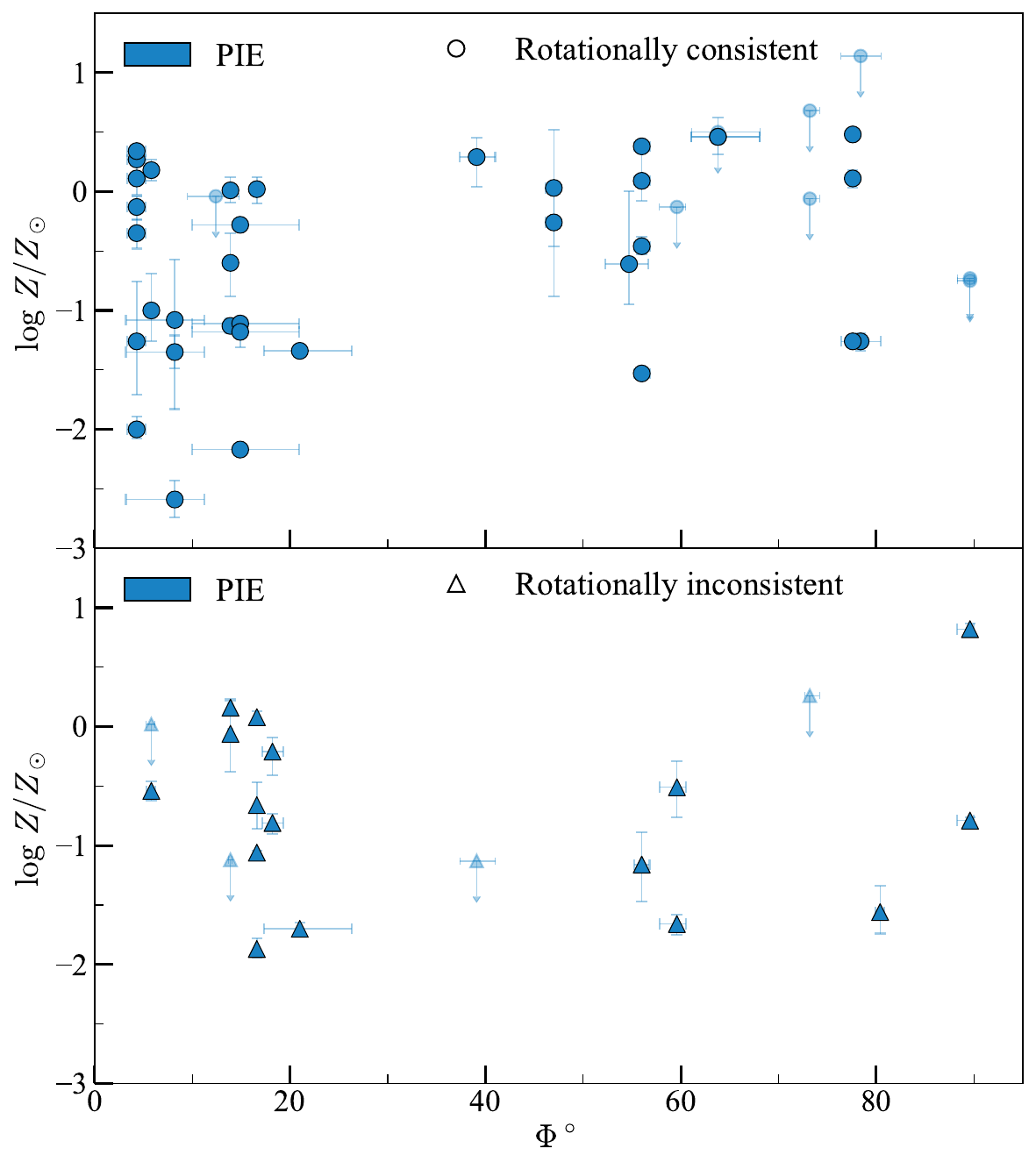}
    \includegraphics[width=0.495\textwidth]{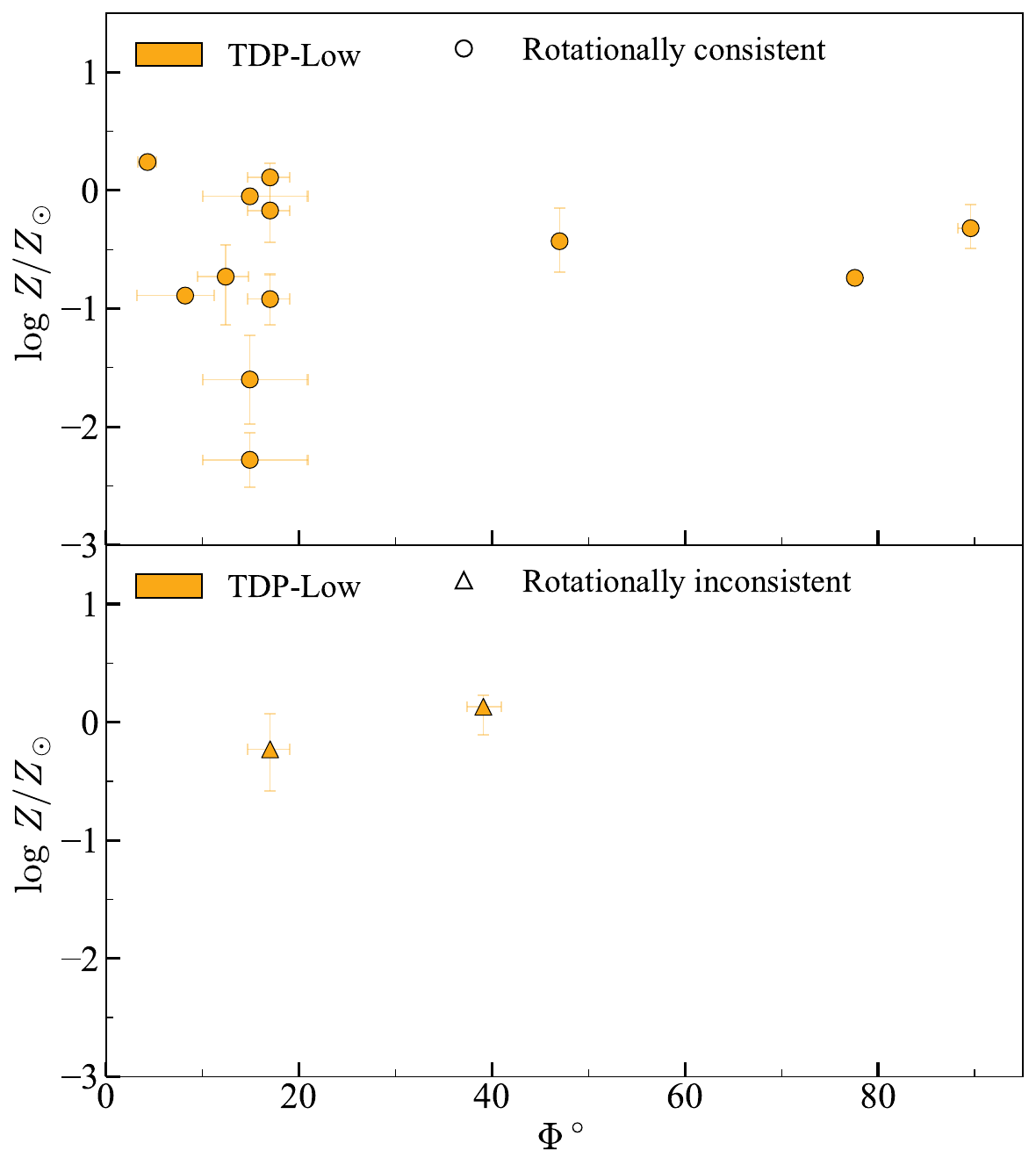}
    \caption{Metallicity of individual CGM clouds as a function of azimuthal angle $\Phi$, separated by ionisation phase and kinematic classification. Left column: photo-ionisation equilibrium (PIE) clouds. Right column: time-dependent photo-ionisation clouds (TDP-low). Top row: clouds whose line-of-sight velocities are consistent with the direction of host galaxy rotation (rotation-consistent). Bottom row: rotation-inconsistent clouds. The azimuthal metallicity dependence is seen for the rotation-consistent PIE population, and hinted at in the TDP-low population, which shows systematically lower metallicities near the major axis ($\Phi<30^\circ$) direction (circles), while the bottom row shows rotationally inconsistent clouds (triangles).
}
    \label{fig:fig1}
\end{figure*}

Figure~\ref{fig:fig1} shows the metallicity of individual clouds as a function of azimuthal angle $\Phi$. The left column shows PIE clouds and the right column shows TDP-low clouds, while the top and bottom rows separate rotation-consistent and rotation-inconsistent populations, respectively. We find 43 PIE and 12 TDP-low rotation-consistent clouds and 20 PIE and 2 TDP-low rotation-inconsistent clouds.

\subsection{Low Ionisation PIE Clouds}
When all clouds are considered together, no clear azimuthal metallicity dependence is apparent \citep{sameer24}. However, an azimuthal metallicity trend emerges within the co-rotating PIE population. In the top-left panel, low-metallicity clouds are preferentially found at small azimuthal angles, with a large fraction of systems at $\Phi \lesssim 30^{\circ}$ exhibiting $\log (Z/Z_{\odot}) \lesssim -1$. In contrast, at larger azimuthal angles the co-rotating PIE clouds are systematically more metal enriched, with few systems below $\log (Z/Z_{\odot}) = -1$. The bootstrap estimate of the mean metallicity difference between the two populations is
$\Delta \langle \log(Z/Z_\odot)\rangle 
= \langle \log Z\rangle_{\Phi<30^\circ} 
- \langle \log Z\rangle_{\Phi>30^\circ}
= -0.52 \pm 0.20$ (2.6$\sigma$). Censored two-sample survival analysis tests do not show a significant difference between the two metallicity distributions ($p=0.18 \pm 0.27$). The large variability in the bootstrapped p-values suggests that the test is likely underpowered for this sample size. Nonetheless, this indicates that the rotation-consistent PIE clouds may differ in mean metallicity with azimuthal angle, even though the full censored distributions overlap.

No comparable azimuthal trend is present in the rotation-inconsistent PIE population ($\Delta \langle \log(Z/Z_\odot)\rangle = 0.13 \pm 0.34$), demonstrating that the azimuthal metallicity signal is strongly suppressed when kinematically distinct gas is included. These rotation-inconsistent clouds span a wide metallicity range at all azimuthal angles, including both low- and moderate-metallicity systems near the projected major and minor axes. Moreover, the mean metallicity difference between rotation-consistent and rotation-inconsistent PIE clouds is consistent with zero ($\Delta \langle \log(Z/Z_\odot)\rangle =0.17 \pm 0.21$), indicating that they are drawn from a similar metallicity distribution.

\begin{figure*}
\centering
\includegraphics[width=0.325\textwidth]{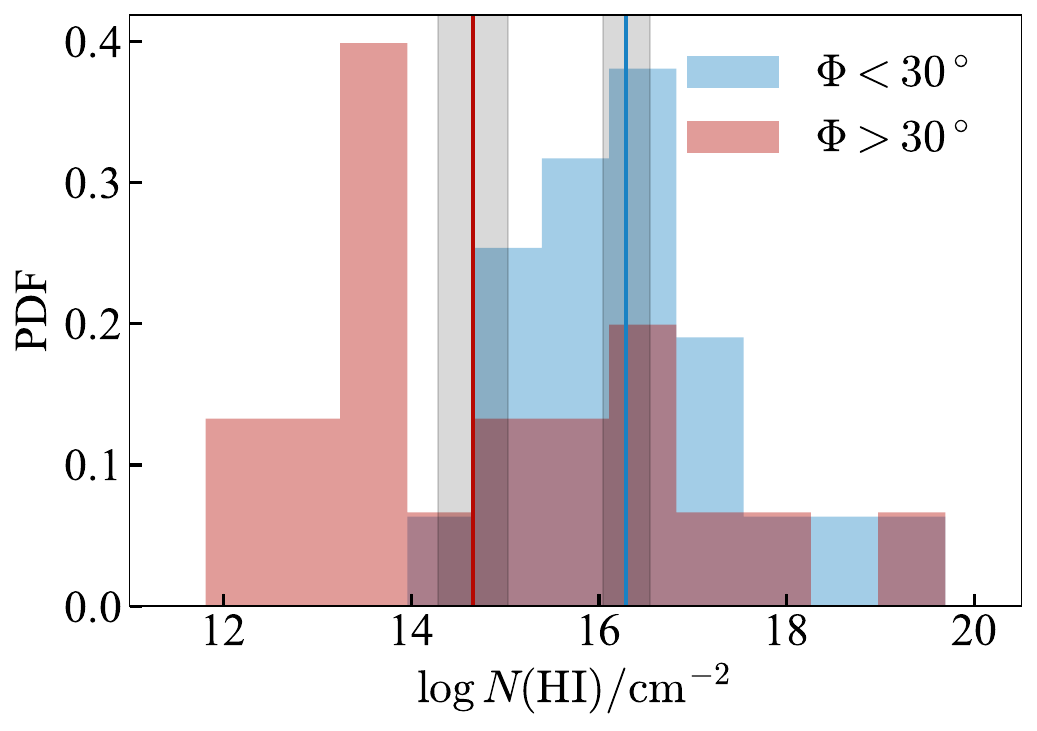}
\includegraphics[width=0.325\textwidth]{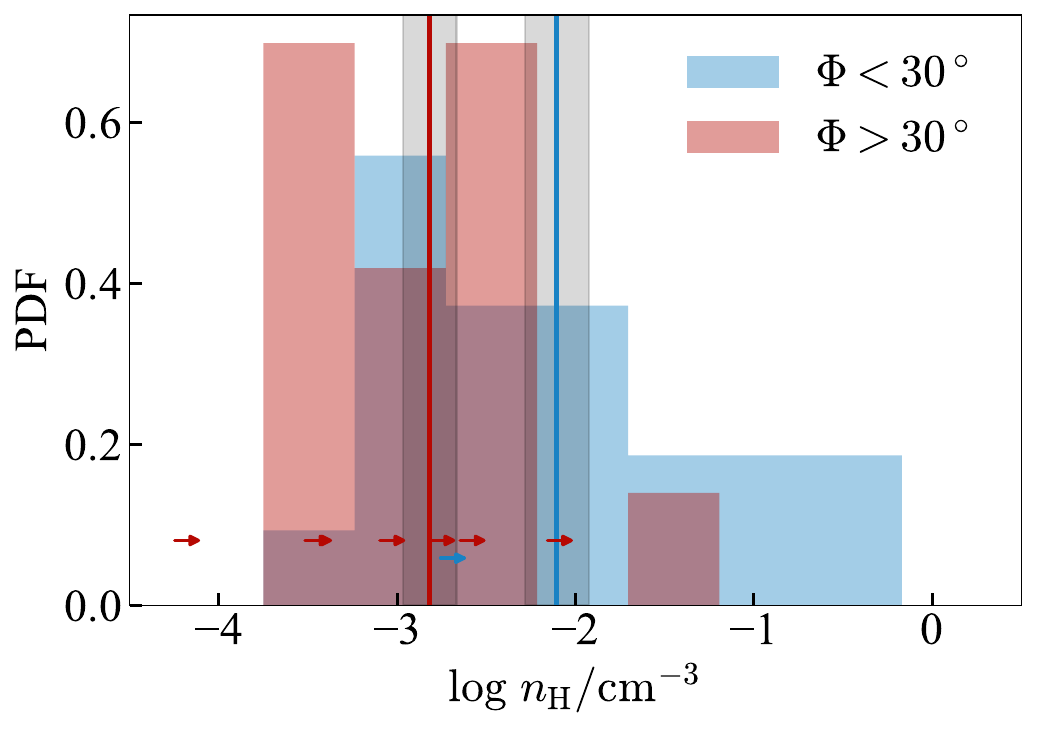}
\includegraphics[width=0.325\textwidth]{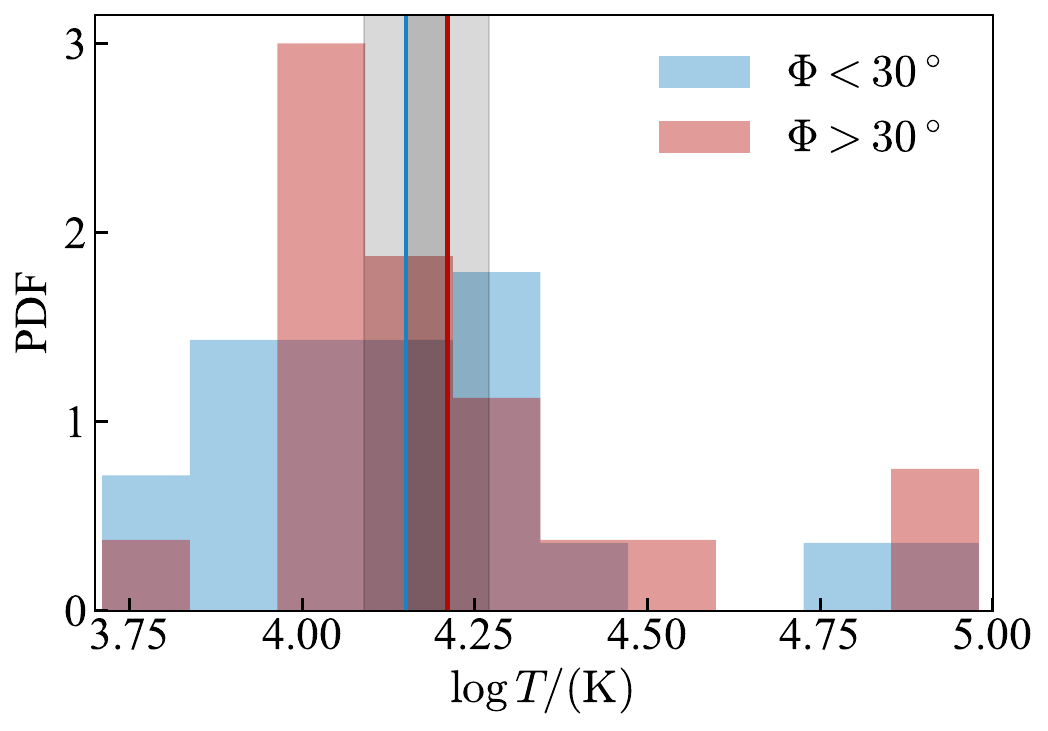}
\includegraphics[width=0.325\textwidth]{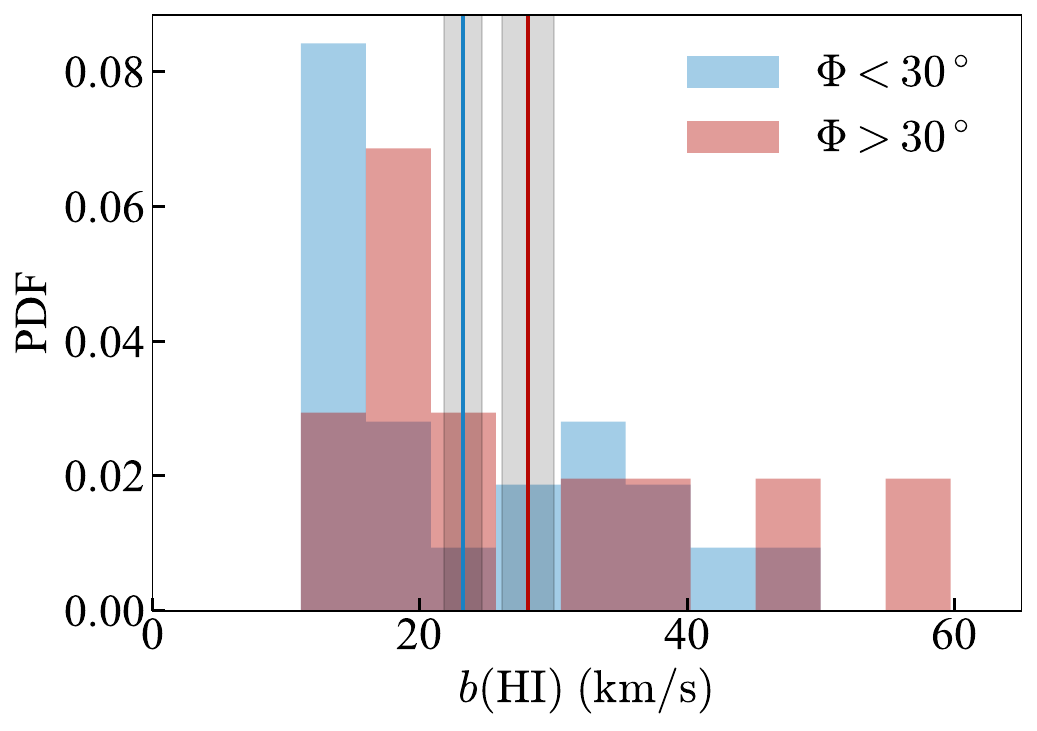}
\includegraphics[width=0.325\textwidth]{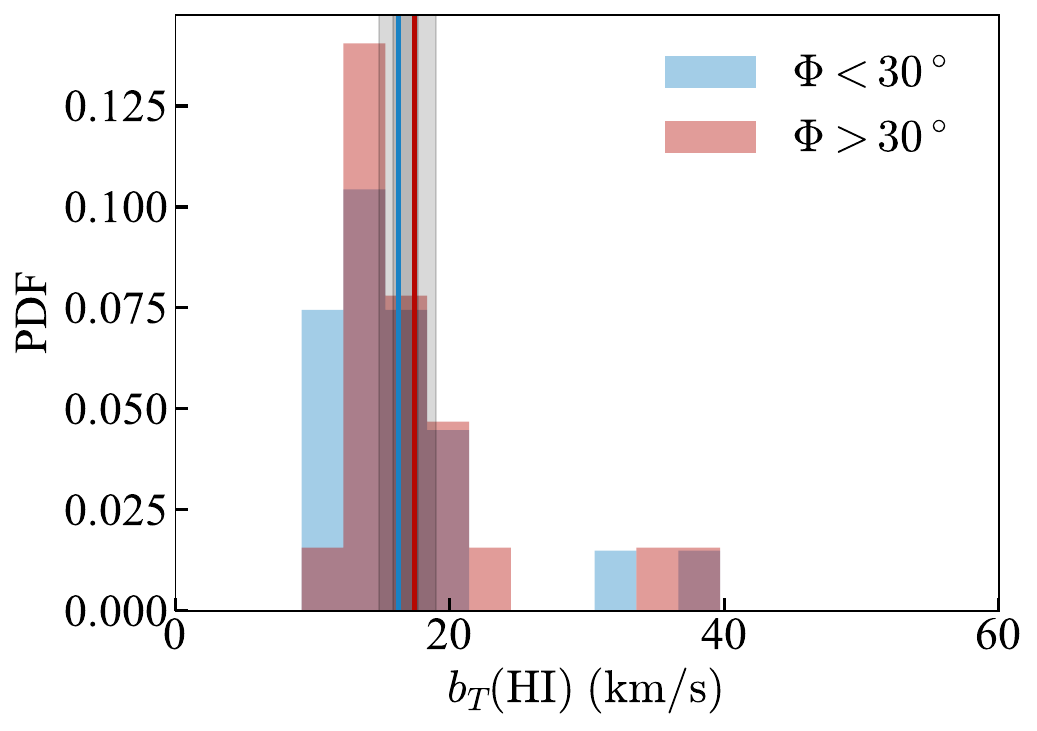}
\includegraphics[width=0.325\textwidth]{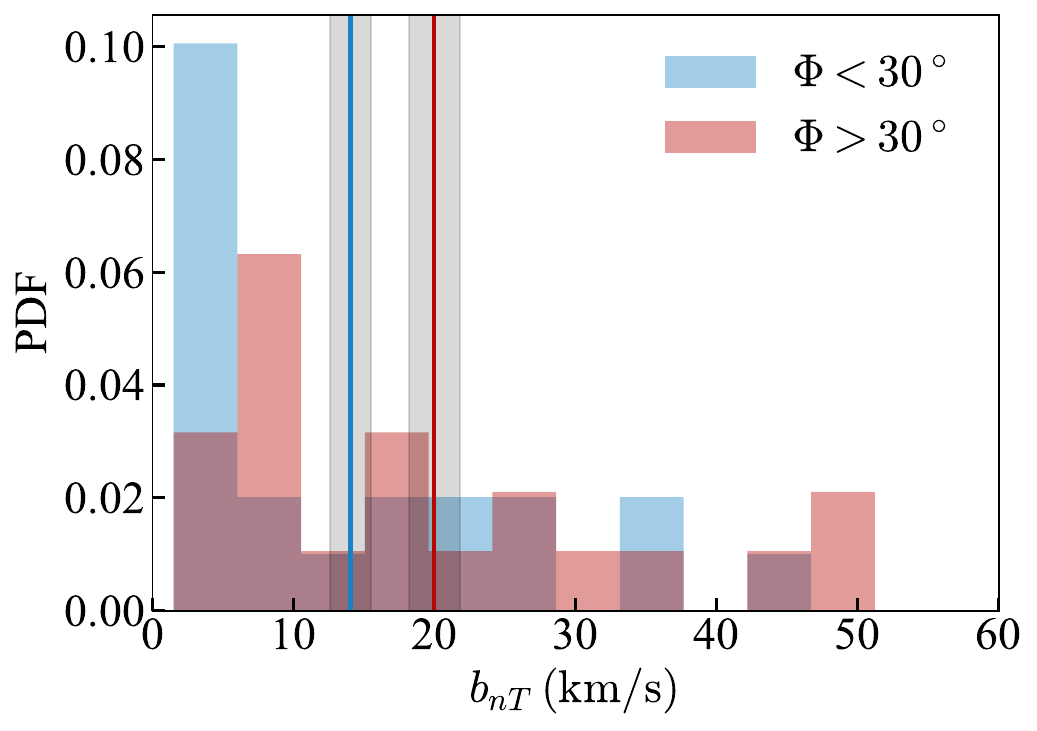}

\caption{Probability distribution functions of the physical parameters for the 
\textit{rotation-consistent PIE} cloud population, split at 
$\Phi_{\rm cut}=30^\circ$. Major-axis dominated clouds 
($\Phi < 30^\circ$; blue) are compared to clouds at larger azimuthal 
angles ($\Phi > 30^\circ$; red). Panels show (from left to right) 
$\log N(\HI)$, $\log n_{\rm H}$, $\log T$, 
$b(\HI)$, the thermal Doppler parameter 
$b_{\rm T}(\HI)$, and the non-thermal component 
$b_{\rm nt}$. The arrows indicate lower limits for $\log n_{\rm H}$. Vertical lines indicate the mean of each 
distribution and the grey shaded region indicates the standard error in the mean. Major-axis clouds exhibit systematically higher $\log N(\HI)$ and $\log n_{\rm H}$, and lower 
$b_{\rm nt}$ compared to clouds at $\Phi > 30^\circ$, while 
$\log T$ and $b_{\rm T}(\HI)$ are statistically consistent (within 1$\sigma$)
between the two azimuthal bins. The absence of a temperature offset 
combined with enhanced non-thermal broadening at larger $\Phi$ 
demonstrates that the azimuthal separation is driven by differences 
in the dynamical state of the cool gas rather than by thermal 
variations.}
\label{fig:fig2}
\end{figure*}

The azimuthal angle metallicity trend for rotation-consistent PIE clouds is further supported by coherent differences in other inferred physical properties (Table~\ref{tab:tab1}; Fig.~\ref{fig:fig2}). Fig.~\ref{fig:fig2} shows the probability distribution functions of the rotation-consistent PIE clouds split at $\Phi_{\rm cut}=30^\circ$. Panels show (from left to right, top to bottom) the cloud {\HI} column density $\log N(\HI)$, hydrogen density $\log n_{\rm H}$, Temperature $\log T$, Doppler parameter $b(\HI)$, the thermal Doppler parameter $b_{\rm T}(\HI)$, and the non-thermal component $b_{\rm nt}$. Vertical lines indicate the mean of each distribution and the grey shaded region indicates the standard error in the mean. 

We find that the $\Phi < 30^\circ$ clouds exhibit systematically higher $\log N(\mathrm{\HI})$ and $\log n_{\rm H}$ relative to the $\Phi > 30^\circ$ population ($\Delta \langle \log N(\mathrm{\HI}) \rangle = +1.63 \pm 0.45$ and $\Delta \langle \log n_{\rm H} \rangle = +0.71 \pm 0.22$), indicating that major-axis co-rotating clouds are denser and contain substantially larger neutral hydrogen column densities. We note that the lower {\HI} column densities at $\Phi>30^\circ$ may raise the metallicity detection floor and reduce sensitivity to the most metal-poor clouds, although this effect alone is unlikely to explain the azimuthal metallicity difference because low-metallicity clouds are still detected at higher azimuthal angles when the kinematic selection is removed (see Figure~\ref{fig:fig1}).

In contrast, the $\log T$ distributions overlap entirely ($\Delta \langle \log T \rangle = -0.06 \pm 0.09$), and the thermal Doppler parameter $b_{\rm T}(\mathrm{\HI})$ is statistically consistent between azimuthal angle bins. The difference in total line width therefore arises from the non-thermal component. The $b_{\rm nT}$ distribution for $\Phi > 30^\circ$ clouds is shifted toward larger velocities ($\Delta \langle b_{\rm nT} \rangle = +5.92 \pm 2.32 \ \mathrm{km\,s^{-1}}$), indicating enhanced turbulence, velocity shear, or unresolved substructure in this population. The total Doppler parameter $b(\mathrm{\HI})$ of the $\Phi > 30^\circ$ clouds is marginally larger ($\Delta \langle b(\mathrm{\HI}) \rangle 
= 4.88 \pm 2.39 \ \mathrm{km\,s^{-1}})$, reflecting the difference seen in the non-thermal Doppler parameter. 

We also find that the differences between rotation-consistent and rotation-inconsistent PIE clouds for all of these properties are consistent with zero, indicating that they are drawn from similar cloud distributions. Similar to \citet{sameer24}, we do not find any additional trends with galaxy properties such as $D/\Rv$, stellar mass, inclination angle, relative galaxy-cloud velocities, etc. We also do not find any evidence that the lower-$N(\mathrm{\HI})$, higher-metallicity co-rotating clouds at larger azimuthal angles can be explained or mitigated by controlling for any of the galaxy properties examined here. 

We further find that the distribution of galaxy properties are consistent between the co-rotating PIE and the non-co-rotating PIE populations. The co-rotating population has mean values for their stellar mass of $\langle \log(M_\star/M_{\odot}) \rangle = 10.28 \pm 0.48$, impact parameter of $\langle D \rangle = 64 \pm 22$~kpc,  virial radius normalised impact parameter of $\langle D/\Rv \rangle = 0.41 \pm 0.25$, and inclination angle of $\langle i \rangle = 59 \pm 12$ degrees. These are consistent with the non-co-rotating
population having mean values for their stellar mass of $\langle \log(M_\star/M_{\odot}) \rangle = 10.42 \pm 0.36$, impact parameter of $\langle D \rangle = 81 \pm 53$~kpc,  virial radius normalised impact parameter of $\langle D/\Rv \rangle = 0.48 \pm 0.25$, and inclination angle of $\langle i \rangle = 59 \pm 16$ degrees.

Taken together, the combination of lower metallicity, higher $N(\mathrm{\HI})$, higher density and reduced non-thermal broadening demonstrates that the $\Phi < 30^\circ$ co-rotating PIE clouds occupy a dynamically colder and chemically distinct regime relative to the co-rotating population at larger azimuthal angles. 

\begin{table}
  \centering
  \caption{Summary statistics for the rotation-consistent PIE subsample split by azimuthal angle ($\Phi$). Values represent the sample mean of each quantity, with uncertainties corresponding to the bootstrap standard error of the mean. The final column lists the difference between the two azimuthal bins ($\Phi<30^\circ - \Phi>30^\circ$) with its propagated uncertainty.}
  \label{tab:consistent_phi_stats}
  \begin{tabular}{lrrc}
    \hline\hline
    Quantity & $\Phi < 30^\circ$ & $\Phi > 30^\circ$ & Difference  \\
         & (consistent) &  (consistent) & ($<30^\circ - >30^\circ$) \\
    \hline
    $\log Z/Z_{\odot}$                   & $-0.77 \pm 0.16$  & $-0.25 \pm 0.13$ & $-0.52 \pm 0.20$ (2.6$\sigma$)\\
    $\log N({\rm HI})/{\rm cm}^{-2}$            & $16.29 \pm 0.25$  & $14.66 \pm 0.37$  & $+1.63 \pm 0.45$ (3.6$\sigma$)\\
      $\log\,n_{\rm H}/{\rm cm}^{-3}$                   & $-2.11 \pm 0.18$   & $-2.82 \pm 0.15$  & $+0.71 \pm 0.22$ (3.2$\sigma$) \\ 
      $\log T/({\rm K})$                           & $4.15 \pm 0.06$   & $4.21 \pm 0.06$   & $-0.06 \pm 0.09$ (0.7$\sigma$)\\    
      $b({\rm HI})\;(\rm{km/s})$            & $23.22 \pm 1.42$   & $28.10 \pm 1.95$   & $-4.88 \pm 2.39$ (2.0$\sigma$)\\
    $b_T({\rm HI})\;(\rm {km/s})$          & $16.27 \pm 1.41$   & $17.42 \pm 1.57$  & $-1.15 \pm 2.11$ (0.5$\sigma$)\\
     $b_{nT}\;(\rm{km/s})$                  & $14.05 \pm 1.44$   & $19.97 \pm 1.81$   & $-5.92 \pm 2.32$ (2.6$\sigma$)\\
  
    \hline
  \end{tabular}
\label{tab:tab1}
\end{table}

\subsection{High/Intermediate-Ionisation TDP-low Clouds}
The higher-ionisation TDP-low phase exhibits a similar metallicity-azimuthal distribution to the PIE clouds. Figure~\ref{fig:fig1} (top-right panel) shows that the $\Phi < 30^\circ$  co-rotating clouds span a range of metallicities, including four clouds with $\log(Z/Z_\odot)\lesssim -1$. The co-rotating clouds with $\Phi > 30^\circ$ have metallicities around $\log(Z/Z_\odot)\lesssim -0.5$.  The bootstrap estimate of the mean metallicity difference between these two populations is
$\Delta \langle \log(Z/Z_\odot)\rangle 
= \langle \log Z\rangle_{\Phi<30^\circ} 
- \langle \log Z\rangle_{\Phi>30^\circ}
= -0.20 \pm 0.27$. The two rotation-inconsistent TDP-low population (bottom-right panel) are metal enriched and lacking low-metallicity clouds. While the TDP-low population does not show a significant azimuthal metallicity dependence, this may reflect the small sample size or indicate that the higher-ionisation gas traces different components of the CGM.

To explore this further, we compare the properties of the co-rotating major-axis PIE and TDP-low clouds. If both phases arise within the same accreting structure traced along the major axis, then they may be expected to share some common physical properties, even if they probe different ionisation states of the gas. We therefore test whether the co-rotating $\Phi < 30^\circ$ TDP-low clouds resemble the co-rotating $\Phi < 30^\circ$ PIE clouds. We find that the $\Phi < 30^\circ$ TDP-low clouds exhibit the following properties: $\log N(\mathrm{\HI})/{\rm cm}^{-2}=14.78\pm0.39$, $\log T/({\rm K})=4.68\pm0.10$,  $b(\mathrm{\HI})=38.19\pm2.51~{\kms}$, $b_{\rm T}(\mathrm{\HI})=28.07\pm2.39~{\kms}$, $b_{\rm nT}=23.87\pm2.50~{\kms}$ and $\log n_{\rm H}/{\rm cm}^{-3}=-3.00\pm0.65$. 

Comparing these values directly to the $\Phi < 30^\circ$ PIE clouds (Table~\ref{tab:tab1}), we find that the $\Phi < 30^\circ$ TDP-low clouds exhibit systematically lower $\log N(\mathrm{\HI})/{\rm cm}^{-2}$ (by $-1.51\pm0.46$), higher $\log T$ ($+0.53\pm0.11$), larger $b(\mathrm{\HI})$ ($+14.98\pm2.99$~{\kms}), larger $b_{\rm T}(\mathrm{\HI})$ ($+11.79\pm2.79$~{\kms}) and larger $b_{\rm nT}$ ($+9.82\pm2.88$~{\kms}), with lower $\log n_{\rm H}/{\rm cm}^{-3}$ ($-0.89\pm0.67$).

Overall, while the co-rotating TDP-low clouds may show a similar metallicity distribution to the co-rotating PIE clouds, the physical properties of the two populations differ significantly. This behaviour is consistent with higher-ionisation gas being less tightly coupled to disk kinematics and more dominated by hot, dynamically complex, low column density or volume filling, material \citep{kacprzak19kine,nateghi24GFII,kacprzak25}.

\section{Discussion}

\subsection{Angular momentum reveals a metallicity bimodality}

When all clouds are considered together, no azimuthal metallicity dependence is detected, consistent with both previous sightline-averaged studies \citep{prochaska17,peroux16,pointon19} and the cloud-by-cloud analysis of \citet{sameer24}. However, once clouds are restricted to those whose velocities are consistent with disk rotation, a metallicity separation emerges in which rotation-consistent PIE clouds are systematically more metal poor along the projected major axis and more enriched toward higher azimuthal angles by $\sim$0.5 dex.

This behaviour demonstrates that geometric selection alone is insufficient to isolate physically distinct gas flows. Theoretical work emphasises that large-scale structure both collimates cold inflows and supplies angular momentum, such that dynamically coherent gas retains memory of its accretion history \citep{fauchergiguere23}. By isolating the angular-momentum-coupled component of the CGM, we reveal chemically distinct sub-populations that are otherwise blended when kinematically inconsistent gas is included.

 The MEGAFLOW survey reported enhanced minor-axis enrichment of up to $\sim1$ dex using [Zn/Fe] as a dust-depletion proxy \citep{wendt21}. Although the amplitude of our metallicity difference is smaller, the qualitative agreement supports the presence of chemically distinct inflow- and feedback-dominated geometries. Differences in tracers, total sightline absorption versus cloud-by-cloud measurements, and sample selection likely account for the quantitative discrepancy. Cosmological simulations predict that while angular inflow–outflow modulation persists across a wide stellar-mass range, the absolute metallicity contrast depends on stellar mass, star-formation activity, redshift, and impact parameter \citep{peroux20}.  Thus, the diversity of host galaxy properties in our sample likely dilutes the amplitude of the metallicity gradient.


\subsection{Coherent inflow along the disk plane}

The azimuthal metallicity separation in the rotation-consistent PIE population is accompanied by coherent shifts in multiple independent physical parameters. Major-axis ($\Phi < 30^\circ$) clouds exhibit higher $N(\mathrm{H\,I})$, higher density, and reduced non-thermal broadening relative to co-rotating clouds at larger azimuthal angles, while temperatures and thermal Doppler parameters are statistically consistent. The absence of a temperature offset demonstrates that the separation is dynamical rather than thermal in origin. The enhanced non-thermal broadening at higher $\Phi$ therefore reflects increased turbulence, velocity shear, or unresolved substructure rather than heating. Together, this suggests that these major-axis PIE clouds trace more spatially coherent structures, whereas the dynamically broader high-$\Phi$ clouds are more consistent with gas experiencing stronger shear or mixing in a disturbed multiphase environment \citep[e.g.][]{magiicat5,gronke18,fielding20,fielding22}.

Taken together, the combination of lower metallicity, higher density, and reduced non-thermal motions is consistent with metal-poor, filamentary inflow that retains substantial angular momentum as it spirals toward the disk \citep{vandevoort11,stewart11,danovich15,ho19,hafen22,stern24}. Simulations predict that cool accreting gas in L$^\ast$ halos at $z<1$ remains dynamically ordered along the disk plane, preserving rotational coherence while remaining chemically less enriched than wind-dominated regions \citep{peroux20,vandevoort21}. The multi-parameter segregation observed here matches this theoretical picture.

Notably, we detect no dependence on $D/R_{\rm vir}$ or other global galaxy properties within this subsample. The absence of a radial trend suggests that the chemodynamical segregation is primarily angular rather than radial (at least within the virial radius), consistent with geometric inflow/outflow predictions from simulations \citep{peroux20}.

Interestingly, there is a significant population of near solar metallicity rotationally consistent clouds along the major axis. These systems may trace the outer regions of extended gaseous disks or recycled accretion that has already settled into the disk plane. Detailed modelling \citep[e.g.,][]{churchill25a,churchill25b} of these individual systems  may help shed light on their physical origin.  

\subsection{{Are enriched high-$\Phi$ clouds outflows or recycled accretion?}}

A classical interpretation of azimuthal CGM bimodality is that minor-axis absorbers trace bipolar winds \citep{bordoloi11,bouche12,kacprzak12,magiicat5}. The higher metallicities and enhanced non-thermal broadening observed at $\Phi \gtrsim 30^\circ$ are qualitatively compatible with wind-driven stirring or wind–halo interfaces. However, these clouds remain kinematically consistent with disk rotation, disfavouring a purely ballistic outflow interpretation. Cosmological simulations show that outflows preferentially emerge along low-density polar channels and rapidly lose coherent rotational support as they propagate \citep{nelson19,muratov17}. 

Observational surveys provide further context. The MEGAFLOW survey reports minor-axis absorbers exhibiting velocity offsets and kinematic asymmetries consistent with large-scale galactic winds \citep{schroetter16,zabl19,schroetter19,wendt21}. In contrast, the high-$\Phi$ PIE clouds in our sample remain kinematically consistent with disk rotation, lacking the large velocity offsets expected for freely expanding winds. This distinction suggests that, although enriched and dynamically complex, the high-azimuthal rotation-consistent clouds more likely trace angular-momentum-supported recycled accretion rather than ballistic outflows. We note that outflows may be present in the clouds that are kinematically inconsistent with galaxy rotation along the minor axis. 

Gas that retains angular-momentum coherence is more naturally explained as recycled accretion in galactic fountains, where enriched material launched in previous outflows cools and reaccretes while conserving a substantial fraction of its angular momentum \citep{angles17,stern24}. Such recycled gas is predicted to be both metal rich and dynamically complex, with enhanced turbulence arising from shear and mixing at the disk–halo interface \citep{hafen19,hafen22,fielding22}. The elevated $b_{\rm nt}$ at higher azimuthal angles without a corresponding temperature increase is consistent with this mixing-driven picture. Simulations further show that recycled and mixing-layer gas can produce kinematically complex absorption with substantial non-thermal broadening as shear-driven turbulence develops in the disk–halo interface \citep[e.g.][]{ticoras26}.

\subsection{Multi-phase chemodynamical structure of the CGM}

The lack of a significant azimuthal metallicity dependence in the TDP-low phase, together with the clear physical differences between the co-rotating PIE and TDP-low major-axis clouds, further reinforces a multi-phase interpretation. While the cool PIE clouds reveal a dynamically cold, metal-poor inflow channel along the disk plane, the warm non-equilibrium gas appears dominated by enriched, dynamically complex, low column density material. This phase stratification aligns with observational evidence that kinematic coupling decreases with increasing ionisation state \citep{nateghi24GFII,kacprzak25} and with simulations in which recycled and outflow-associated gas preferentially populates warmer phases before cooling \citep{muratov17,marinacci19,hafen22,decataldo24}. Recent high-resolution simulations similarly show that different ions probe distinct dynamical structures in the CGM. In IllustrisTNG simulations, cool gas is substantially more aligned with the disk angular momentum axis than warm gas, and the corotating fraction declines from low ions such as {\MgII} toward higher-ionisation species such as {\OVI} \citep{messere26}. In the FOGGIE simulations, low-ionisation absorbers such as {\MgII} arise in cooler, spatially coherent structures that remain dynamically coupled to galactic flows, while higher-ionisation species such as {\OVI} preferentially trace more diffuse, kinematically complex gas \citep{ticoras26}. This phase-dependent kinematic behaviour supports the interpretation that the cool PIE population isolates dynamically coherent inflow, whereas warmer phases increasingly trace recycled and mixing-layer material. These results emphasise that metallicity gradients cannot be interpreted independently of ionisation structure and kinematic coupling, as different phases sample distinct regions of the baryon cycle.

\section*{Conclusions}

We demonstrate that an azimuthal metallicity gradient in the cool CGM becomes apparent only after isolating clouds that are dynamically coupled to disk rotation. Rotation-consistent PIE clouds along the projected major axis are systematically more metal-poor, denser and exhibit reduced non-thermal broadening compared to co-rotating clouds at larger azimuthal angles. The absence of any temperature offset indicates that the separation is dynamical rather than thermal in origin. These multi-parameter differences are consistent with dynamically coherent, metal-poor inflow along the disk plane and enriched, more turbulent recycled accretion at higher azimuth that retains substantial angular momentum.

While the higher-ionisation TDP-low phase exhibits a broadly similar azimuthal metallicity distribution to the PIE clouds, it does not show a significant metallicity dependence on azimuthal angle. However, major-axis PIE and TDP-low co-rotating clouds do not exhibit similar physical properties, with the TDP-low clouds being hotter, more dynamically complex, lower in column density and lower in density. This suggests that the higher-ionisation phase is less tightly coupled to the low-ionisation phase and likely traces different components of the CGM.

Our results demonstrate that angular momentum selection is essential for revealing the chemodynamical structure of the CGM. Combining metallicity, density and cloud-kinematics provides a powerful observational pathway to disentangle inflow, recycled accretion, and outflows in galaxy halos.

\section*{Acknowledgements}

We thank the anonymous referee for their constructive and useful comments that improved this manuscript. Some of the data presented herein were obtained at the W. M. Keck Observatory, which is operated as a scientific partnership among the California Institute of Technology, the University of California, and the National Aeronautics and Space Administration. The Observatory was made possible by the generous financial support of the W. M. Keck Foundation. This work is based on data obtained through Swinburne with Keck programs 2010A\_W007E, 2010B\_W032E, 2014A\_W178E, 2014B\_W018E, 2015\_W187E, and 2016A\_W056E. The authors wish to recognise and acknowledge the very significant cultural role and reverence that the summit of Maunakea has always had within the indigenous Hawaiian community. We are most fortunate to have the opportunity to conduct observations from this mountain.

\section*{Data Availability}
The data underlying this paper will be shared on reasonable request to the corresponding author.



\bibliographystyle{mnras}
\bibliography{example}

@ARTICLE{angles17,
       author = {{Angl{\'e}s-Alc{\'a}zar}, Daniel and
                  {Faucher-Gigu{\`e}re}, Claude-Andr{\'e} and
                  {Kere{\v{s}}}, Du{\v{s}}an and {Hopkins}, Philip
                  F. and {Quataert}, Eliot and {Murray}, Norman},
        title = "{The cosmic baryon cycle and galaxy mass assembly in
                  the FIRE simulations}",
      journal = {\mnras},
         year = "2017",
        month = "Oct",
       volume = {470},
       number = {4},
        pages = {4698-4719},
          doi = {10.1093/mnras/stx1517},
archivePrefix = {arXiv},
       eprint = {1610.08523},
 primaryClass = {astro-ph.GA},
       adsurl = {https://ui.adsabs.harvard.edu/abs/2017MNRAS.470.4698A}
}

@ARTICLE{beckett21,
       author = {{Beckett}, Alexander and {Morris}, Simon L. and {Fumagalli}, Michele and {Bielby}, Rich and {Tejos}, Nicolas and {Schaye}, Joop and {Jannuzi}, Buell and {Cantalupo}, Sebastiano},
        title = "{The relationship between gas and galaxies at z < 1 using the Q0107 quasar triplet}",
      journal = {\mnras},
         year = 2021,
        month = sep,
       volume = {506},
       number = {2},
        pages = {2574-2602},
          doi = {10.1093/mnras/stab1630},
archivePrefix = {arXiv},
       eprint = {2106.06416},
 primaryClass = {astro-ph.GA},
       adsurl = {https://ui.adsabs.harvard.edu/abs/2021MNRAS.506.2574B}
}

@ARTICLE{bordoloi11,
   author = {{Bordoloi}, R. and {Lilly}, S.~J. and {Knobel}, C. and
                  {Bolzonella}, M. and {Kampczyk}, P. and {Carollo},
                  C.~M. and {Iovino}, A. and {Zucca}, E. and
                  {Contini}, T. and {Kneib}, J.-P. and {Le Fevre},
                  O. and {Mainieri}, V. and {Renzini}, A. and
                  {Scodeggio}, M. and {Zamorani}, G. and {Balestra},
                  I. and {Bardelli}, S. and {Bongiorno}, A. and
                  {Caputi}, K. and {Cucciati}, O. and {de la Torre},
                  S. and {de Ravel}, L. and {Garilli}, B. and {Kova{\v
                  c}}, K. and {Lamareille}, F. and {Le Borgne},
                  J.-F. and {Le Brun}, V. and {Maier}, C. and
                  {Mignoli}, M. and {Pello}, R. and {Peng}, Y. and
                  {Perez Montero}, E. and {Presotto}, V. and
                  {Scarlata}, C. and {Silverman}, J. and {Tanaka},
                  M. and {Tasca}, L. and {Tresse}, L. and {Vergani},
                  D. and {Barnes}, L. and {Cappi}, A. and {Cimatti},
                  A. and {Coppa}, G. and {Diener}, C. and {Franzetti},
                  P. and {Koekemoer}, A. and {L{\'o}pez-Sanjuan},
                  C. and {McCracken}, H.~J. and {Moresco}, M. and
                  {Nair}, P. and {Oesch}, P. and {Pozzetti}, L. and
                  {Welikala}, N.  },
    title = "{The Radial and Azimuthal Profiles of Mg II Absorption
                  around 0.5 {\lt} z {\lt} 0.9 zCOSMOS Galaxies of
                  Different Colors, Masses, and Environments}",
  journal = {\apj},
archivePrefix = "arXiv",
   eprint = {1106.0616},
 primaryClass = "astro-ph.CO",
     year = 2011,
    month = dec,
   volume = 743,
      eid = {10},
    pages = {10},
      doi = {10.1088/0004-637X/743/1/10},
   adsurl = {http://ui.adsabs.harvard.edu/abs/2011ApJ...743...10B}
}

@ARTICLE{bouche12,
   author = {{Bouch{\'e}}, N. and {Hohensee}, W. and {Vargas}, R. and
                  {Kacprzak}, G.~G. and {Martin}, C.~L. and {Cooke},
                  J. and {Churchill}, C.~W.},
    title = "{Physical properties of galactic winds using background
                  quasars}",
  journal = {\mnras},
archivePrefix = "arXiv",
   eprint = {1110.5877},
 primaryClass = "astro-ph.CO",
     year = 2012,
    month = oct,
   volume = 426,
    pages = {801-815},
      doi = {10.1111/j.1365-2966.2012.21114.x},
   adsurl = {http://ui.adsabs.harvard.edu/abs/2012MNRAS.426..801B}
}

@ARTICLE{bouche13,
   author = {{Bouch{\'e}}, N. and {Murphy}, M.~T. and {Kacprzak},
                  G.~G. and {P{\'e}roux}, C. and {Contini}, T. and
                  {Martin}, C.~L. and {Dessauges-Zavadsky}, M.  },
    title = "{Signatures of Cool Gas Fueling a Star-Forming Galaxy at
                  Redshift 2.3}",
  journal = {Science},
archivePrefix = "arXiv",
   eprint = {1306.0134},
 primaryClass = "astro-ph.CO",
     year = 2013,
    month = jul,
   volume = 341,
    pages = {50-53},
      doi = {10.1126/science.1234209},
   adsurl = {http://ui.adsabs.harvard.edu/abs/2013Sci...341...50B}
}

@ARTICLE{bouche16,
       author = {{Bouch{\'e}}, N. and {Finley}, H. and {Schroetter}, I. and {Murphy}, M.~T. and {Richter}, P. and {Bacon}, R. and {Contini}, T. and {Richard}, J. and {Wendt}, M. and {Kamann}, S. and {Epinat}, B. and {Cantalupo}, S. and {Straka}, L.~A. and {Schaye}, J. and {Martin}, C.~L. and {P{\'e}roux}, C. and {Wisotzki}, L. and {Soto}, K. and {Lilly}, S. and {Carollo}, C.~M. and {Brinchmann}, J. and {Kollatschny}, W.},
        title = "{Possible Signatures of a Cold-flow Disk from MUSE Using a z {\ensuremath{\sim}} 1 Galaxy-Quasar Pair toward SDSS J1422-0001}",
      journal = {\apj},
         year = 2016,
        month = apr,
       volume = {820},
       number = {2},
          eid = {121},
        pages = {121},
          doi = {10.3847/0004-637X/820/2/121},
archivePrefix = {arXiv},
       eprint = {1601.07567},
 primaryClass = {astro-ph.GA},
       adsurl = {https://ui.adsabs.harvard.edu/abs/2016ApJ...820..121B}
}

@ARTICLE{churchill15,
   author = {{Churchill}, C.~W. and {Vander Vliet}, J.~R. and
                  {Trujillo-Gomez}, S. and {Kacprzak}, G.~G. and
                  {Klypin}, A.},
    title = "{Direct Insights Into Observational Absorption Line
                  Analysis Methods of the Circumgalactic Medium Using
                  Cosmological Simulations}",
  journal = {\apj},
archivePrefix = "arXiv",
   eprint = {1409.0914},
     year = 2015,
    month = mar,
   volume = 802,
      eid = {10},
    pages = {10},
      doi = {10.1088/0004-637X/802/1/10},
   adsurl = {http://ui.adsabs.harvard.edu/abs/2015ApJ...802...10C}
}

@ARTICLE{churchill25a,
       author = {{Churchill}, Christopher W.},
        title = "{Spatial-Kinematic Absorption Models of the Circumgalactic Medium. I. Structures, Orientations, and Kinematics}",
      journal = {arXiv e-prints},
         year = 2025,
        month = oct,
          eid = {arXiv:2510.24357},
        pages = {arXiv:2510.24357},
          doi = {10.48550/arXiv.2510.24357},
archivePrefix = {arXiv},
       eprint = {2510.24357},
 primaryClass = {astro-ph.GA},
       adsurl = {https://ui.adsabs.harvard.edu/abs/2025arXiv251024357C}
}

@ARTICLE{churchill25b,
       author = {{Churchill}, Christopher W.},
        title = "{Spatial-Kinematic Absorption Models of the Circumgalactic Medium. II. Ionized Gas Phases and Absorption Lines}",
      journal = {arXiv e-prints},
         year = 2025,
        month = oct,
          eid = {arXiv:2510.23803},
        pages = {arXiv:2510.23803},
          doi = {10.48550/arXiv.2510.23803},
archivePrefix = {arXiv},
       eprint = {2510.23803},
 primaryClass = {astro-ph.GA},
       adsurl = {https://ui.adsabs.harvard.edu/abs/2025arXiv251023803C}
}

@ARTICLE{danovich15,
   author = {{Danovich}, M. and {Dekel}, A. and {Hahn}, O. and
                  {Ceverino}, D. and {Primack}, J.},
    title = "{Four phases of angular-momentum buildup in high-z
                  galaxies: from cosmic-web streams through an
                  extended ring to disc and bulge}",
  journal = {\mnras},
archivePrefix = "arXiv",
   eprint = {1407.7129},
     year = 2015,
    month = may,
   volume = 449,
    pages = {2087-2111},
      doi = {10.1093/mnras/stv270},
   adsurl = {http://ui.adsabs.harvard.edu/abs/2015MNRAS.449.2087D}
}

@ARTICLE{decataldo24,
       author = {{Decataldo}, Davide and {Shen}, Sijing and {Mayer}, Lucio and {Baumschlager}, Bernhard and {Madau}, Piero},
        title = "{The origin of cold gas in the circumgalactic medium}",
      journal = {\aap},
         year = 2024,
        month = may,
       volume = {685},
          eid = {A8},
        pages = {A8},
          doi = {10.1051/0004-6361/202346972},
archivePrefix = {arXiv},
       eprint = {2306.03146},
 primaryClass = {astro-ph.GA},
       adsurl = {https://ui.adsabs.harvard.edu/abs/2024A&A...685A...8D}
}

@INPROCEEDINGS{dekker-uves,
   author = {{Dekker}, H. and {D'Odorico}, S. and {Kaufer}, A. and
                  {Delabre}, B. and {Kotzlowski}, H.},
    title = "{Design, construction, and performance of UVES, the
                  echelle spectrograph for the UT2 Kueyen Telescope at
                  the ESO Paranal Observatory}",
booktitle = {Optical and IR Telescope Instrumentation and Detectors},
     year = 2000,
   series = {SPIE Conference Series},
   volume = 4008,
   editor = {{Iye}, M. and {Moorwood}, A.~F.},
    month = aug,
    pages = {534-545},
   adsurl = {http://ui.adsabs.harvard.edu/abs/2000SPIE.4008..534D}
}

@ARTICLE{diamond-stanic16,
       author = {{Diamond-Stanic}, Aleksandar M. and {Coil}, Alison L. and {Moustakas}, John and {Tremonti}, Christy A. and {Sell}, Paul H. and {Mendez}, Alexander J. and {Hickox}, Ryan C. and {Rudnick}, Greg H.},
        title = "{Galaxies Probing Galaxies at High Resolution: Co-rotating Gas Associated with a Milky Way Analog at z=0.4}",
      journal = {\apj},
         year = 2016,
        month = jun,
       volume = {824},
       number = {1},
          eid = {24},
        pages = {24},
          doi = {10.3847/0004-637X/824/1/24},
archivePrefix = {arXiv},
       eprint = {1507.01945},
 primaryClass = {astro-ph.GA},
       adsurl = {https://ui.adsabs.harvard.edu/abs/2016ApJ...824...24D}
}

@ARTICLE{dutta26,
       author = {{Dutta}, Sayak and {Muzahid}, Sowgat and {Schaye}, Joop and {Johnson}, Sean and {Herenz}, Edmund Christian and {Pessa}, Ismael and {Augustin}, Ramona and {Bouch{\'e}}, Nicolas F. and {Braspenning}, Joey and {Cantalupo}, Sebastiano and {Das}, Sourav and {Wendt}, Martin},
        title = "{MUSEQuBES: Probing Anisotropies in Gas and Metal Distributions in the Circumgalactic Medium}",
      journal = {arXiv e-prints},
         year = 2026,
        month = feb,
          eid = {arXiv:2602.17593},
        pages = {arXiv:2602.17593},
          doi = {10.48550/arXiv.2602.17593},
archivePrefix = {arXiv},
       eprint = {2602.17593},
 primaryClass = {astro-ph.GA},
       adsurl = {https://ui.adsabs.harvard.edu/abs/2026arXiv260217593D}
}

@ARTICLE{fauchergiguere23,
       author = {{Faucher-Gigu{\`e}re}, Claude-Andr{\'e} and {Oh}, S. Peng},
        title = "{Key Physical Processes in the Circumgalactic Medium}",
      journal = {\araa},
         year = 2023,
        month = aug,
       volume = {61},
        pages = {131-195},
          doi = {10.1146/annurev-astro-052920-125203},
archivePrefix = {arXiv},
       eprint = {2301.10253},
 primaryClass = {astro-ph.GA},
       adsurl = {https://ui.adsabs.harvard.edu/abs/2023ARA&A..61..131F}
}

@ARTICLE{cloudy,
       author = {{Ferland}, G.~J. and {Porter}, R.~L. and {van Hoof},
                  P.~A.~M. and {Williams}, R.~J.~R. and {Abel},
                  N.~P. and {Lykins}, M.~L. and {Shaw}, G. and
                  {Henney}, W.~J. and {Stancil}, P.~C.},
        title = "{The 2013 Release of Cloudy}",
      journal = {\rmxaa},
         year = "2013",
        month = "Apr",
       volume = {49},
        pages = {137-163},
archivePrefix = {arXiv},
       eprint = {1302.4485},
 primaryClass = {astro-ph.GA},
       adsurl = {https://ui.adsabs.harvard.edu/\#abs/2013RMxAA..49..137F}
}

@ARTICLE{green12,
       author = {{Green}, James C. and {Froning}, Cynthia S. and {Osterman}, Steve and {Ebbets}, Dennis and {Heap}, Sara H. and {Leitherer}, Claus and {Linsky}, Jeffrey L. and {Savage}, Blair D. and {Sembach}, Kenneth and {Shull}, J. Michael and {Siegmund}, Oswald H.~W. and {Snow}, Theodore P. and {Spencer}, John and {Stern}, S. Alan and {Stocke}, John and {Welsh}, Barry and {B{\'e}land}, St{\'e}phane and {Burgh}, Eric B. and {Danforth}, Charles and {France}, Kevin and {Keeney}, Brian and {McPhate}, Jason and {Penton}, Steven V. and {Andrews}, John and {Brownsberger}, Kenneth and {Morse}, Jon and {Wilkinson}, Erik},
        title = "{The Cosmic Origins Spectrograph}",
      journal = {\apj},
         year = 2012,
        month = jan,
       volume = {744},
       number = {1},
          eid = {60},
        pages = {60},
          doi = {10.1088/0004-637X/744/1/6010.1086/141956},
archivePrefix = {arXiv},
       eprint = {1110.0462},
 primaryClass = {astro-ph.IM},
       adsurl = {https://ui.adsabs.harvard.edu/abs/2012ApJ...744...60G}
}

@ARTICLE{gronke18,
       author = {{Gronke}, Max and {Oh}, S. Peng},
        title = "{The growth and entrainment of cold gas in a hot wind}",
      journal = {\mnras},
         year = 2018,
        month = oct,
       volume = {480},
       number = {1},
        pages = {L111-L115},
          doi = {10.1093/mnrasl/sly131},
archivePrefix = {arXiv},
       eprint = {1806.02728},
 primaryClass = {astro-ph.GA},
       adsurl = {https://ui.adsabs.harvard.edu/abs/2018MNRAS.480L.111G}
}

@ARTICLE{hafen19,
       author = {{Hafen}, Zachary and {Faucher-Gigu{\`e}re},
                  Claude-Andr{\'e} and {Angl{\'e}s-Alc{\'a}zar},
                  Daniel and {Stern}, Jonathan and {Kere{\v{s}}},
                  Du{\v{s}}an and {Hummels}, Cameron and {Esmerian},
                  Clarke and {Garrison-Kimmel}, Shea and {El-Badry},
                  Kareem and {Wetzel}, Andrew and {Chan}, T.~K. and
                  {Hopkins}, Philip F. and {Murray}, Norman},
        title = "{The origins of the circumgalactic medium in the FIRE
                  simulations}",
      journal = {\mnras},
         year = "2019",
        month = "Sep",
       volume = {488},
       number = {1},
        pages = {1248-1272},
          doi = {10.1093/mnras/stz1773},
archivePrefix = {arXiv},
       eprint = {1811.11753},
 primaryClass = {astro-ph.GA},
       adsurl = {https://ui.adsabs.harvard.edu/abs/2019MNRAS.488.1248H}
}

@ARTICLE{ho26,
       author = {{Ho}, Stephanie H. and {Martin}, Crystal L. and {Nateghi}, Hasti and {Kacprzak}, Glenn G. and {Stern}, Jonathan},
        title = "{Kinematics of Circumgalactic O VI Gas and Disk Rotation of z ≍ 0.2 Star-forming Galaxies}",
      journal = {\apj},
         year = 2026,
        month = feb,
       volume = {998},
       number = {2},
          eid = {261},
        pages = {261},
          doi = {10.3847/1538-4357/ae1b88},
archivePrefix = {arXiv},
       eprint = {2507.11664},
 primaryClass = {astro-ph.GA},
       adsurl = {https://ui.adsabs.harvard.edu/abs/2026ApJ...998..261H}
}

@ARTICLE{ho17,
       author = {{Ho}, Stephanie H. and {Martin}, Crystal L. and
                  {Kacprzak}, Glenn G. and {Churchill}, Christopher
                  W.},
        title = "{Quasars Probing Galaxies. I. Signatures of Gas
                  Accretion at Redshift Approximately 0.2}",
      journal = {\apj},
         year = "2017",
        month = "Feb",
       volume = {835},
       number = {2},
          eid = {267},
        pages = {267},
          doi = {10.3847/1538-4357/835/2/267},
archivePrefix = {arXiv},
       eprint = {1611.04579},
 primaryClass = {astro-ph.GA},
       adsurl = {https://ui.adsabs.harvard.edu/abs/2017ApJ...835..267H}
}

@ARTICLE{ho19,
       author = {{Ho}, Stephanie H. and {Martin}, Crystal L. and {Turner}, Monica L.},
        title = "{How Gas Accretion Feeds Galactic Disks}",
      journal = {\apj},
         year = 2019,
        month = apr,
       volume = {875},
       number = {1},
          eid = {54},
        pages = {54},
          doi = {10.3847/1538-4357/ab0ec2},
archivePrefix = {arXiv},
       eprint = {1903.06840},
 primaryClass = {astro-ph.GA},
       adsurl = {https://ui.adsabs.harvard.edu/abs/2019ApJ...875...54H}
}

@ARTICLE{kacprzak11kin,
   author = {{Kacprzak}, G.~G. and {Churchill}, C.~W. and {Barton},
                  E.~J. and {Cooke}, J.},
    title = "{Halo Gas and Galaxy Disk Kinematics of a Volume-limited
                  Sample of Mg II Absorption-selected Galaxies at z
                  \~{} 0.1}",
  journal = {\apj},
archivePrefix = "arXiv",
   eprint = {1102.4339},
 primaryClass = "astro-ph.CO",
     year = 2011,
    month = jun,
   volume = 733,
      eid = {105},
    pages = {105},
      doi = {10.1088/0004-637X/733/2/105},
   adsurl = {http://ui.adsabs.harvard.edu/abs/2011ApJ...733..105K}
}

@ARTICLE{kacprzak10,
   author = {{Kacprzak}, G.~G. and {Churchill}, C.~W. and {Ceverino},
                  D. and {Steidel}, C.~C. and {Klypin}, A. and
                  {Murphy}, M.~T.},
    title = "{Halo Gas and Galaxy Disk Kinematics Derived from
                  Observations and {$\Lambda$}CDM Simulations of Mg II
                  Absorption-selected Galaxies at Intermediate
                  Redshift}",
  journal = {\apj},
archivePrefix = "arXiv",
   eprint = {0912.2746},
     year = 2010,
    month = mar,
   volume = 711,
    pages = {533-558},
      doi = {10.1088/0004-637X/711/2/533},
   adsurl = {http://ui.adsabs.harvard.edu/abs/2010ApJ...711..533K}
}

@ARTICLE{kacprzak12,
   author = {{Kacprzak}, G.~G. and {Churchill}, C.~W. and {Nielsen},
                  N.~M.},
    title = "{Tracing Outflows and Accretion: A Bimodal Azimuthal
                  Dependence of Mg II Absorption}",
  journal = {\apjl},
archivePrefix = "arXiv",
   eprint = {1205.0245},
 primaryClass = "astro-ph.CO",
     year = 2012,
    month = nov,
   volume = 760,
      eid = {L7},
    pages = {L7},
      doi = {10.1088/2041-8205/760/1/L7},
   adsurl = {http://ui.adsabs.harvard.edu/abs/2012ApJ...760L...7K}
}

@ARTICLE{kacprzak15,
   author = {{Kacprzak}, G.~G. and {Muzahid}, S. and {Churchill},
                  C.~W. and {Nielsen}, N.~M. and {Charlton}, J.~C.},
    title = "{The Azimuthal Dependence of Outflows and Accretion
                  Detected Using O VI Absorption}",
  journal = {\apj},
archivePrefix = "arXiv",
   eprint = {1511.03275},
     year = 2015,
    month = dec,
   volume = 815,
      eid = {22},
    pages = {22},
      doi = {10.1088/0004-637X/815/1/22},
   adsurl = {http://ui.adsabs.harvard.edu/abs/2015ApJ...815...22K}
}

@ARTICLE{kacprzak19kine,
       author = {{Kacprzak}, Glenn G. and {Vander Vliet}, Jacob R. and
                  {Nielsen}, Nikole M. and {Muzahid}, Sowgat and
                  {Pointon}, Stephanie K. and {Churchill}, Christopher
                  W. and {Ceverino}, Daniel and {Arraki}, Kenz S. and
                  {Klypin}, Anatoly and {Charlton}, Jane C. and
                  {Lewis}, James},
        title = "{The Relation between Galaxy ISM and Circumgalactic O
                  VI Gas Kinematics Derived from Observations and
                  {\ensuremath{\Lambda}}CDM Simulations}",
      journal = {\apj},
         year = "2019",
        month = "Jan",
       volume = {870},
          eid = {137},
        pages = {137},
          doi = {10.3847/1538-4357/aaf1a6},
archivePrefix = {arXiv},
       eprint = {1811.06028},
 primaryClass = {astro-ph.GA},
       adsurl = {https://ui.adsabs.harvard.edu/\#abs/2019ApJ...870..137K}
}

@ARTICLE{kacprzak19metals,
       author = {{Kacprzak}, Glenn G. and {Pointon}, Stephanie K. and
                  {Nielsen}, Nikole M. and {Churchill}, Christopher
                  W. and {Muzahid}, Sowgat and {Charlton}, Jane C.},
        title = "{The Relationship between Galaxy ISM and
                  Circumgalactic Gas Metallicities}",
      journal = {\apj},
         year = "2019",
        month = "Dec",
       volume = {886},
       number = {2},
          eid = {91},
        pages = {91},
          doi = {10.3847/1538-4357/ab4c3c},
archivePrefix = {arXiv},
       eprint = {1910.04310},
 primaryClass = {astro-ph.GA},
       adsurl = {https://ui.adsabs.harvard.edu/abs/2019ApJ...886...91K}
}

@ARTICLE{kacprzak25,
       author = {{Kacprzak}, Glenn G. and {Oppenheimer}, Benjamin and {Nielsen}, Nikole and {Fern{\'a}ndez-Figueroa}, Antonia and {Murphy}, Michael T. and {Allen}, Rebecca and {Barone}, Tania and {Sameer}, Sameer and {Churchill}, Christopher W. and {Burchett}, Joseph and {Gupta}, Kaustubh R. and {Charlton}, Jane C. and {Platukis}, Caleb},
        title = "{COS-EDGES: Co-rotation and kinematic stratification of the multi-phase CGM around edge-on galaxies}",
      journal = {\pasa},
         year = 2025,
        month = sep,
       volume = {42},
          eid = {e128},
        pages = {e128},
          doi = {10.1017/pasa.2025.10091},
archivePrefix = {arXiv},
       eprint = {2507.11613},
 primaryClass = {astro-ph.GA},
       adsurl = {https://ui.adsabs.harvard.edu/abs/2025PASA...42..128K}
}

@article{kaplan58,
author = {E. L. Kaplan and Paul Meier},
title = {Nonparametric Estimation from Incomplete Observations},
journal = {Journal of the American Statistical Association},
volume = {53},
number = {282},
pages = {457--481},
year = {1958},
publisher = {Taylor \& Francis},
doi = {10.1080/01621459.1958.10501452},
URL ={https://doi.org/10.1080/01621459.1958.10501452
},
eprint = { 
https://doi.org/10.1080/01621459.1958.10501452
}
}

@ARTICLE{lan14,
   author = {{Lan}, T.-W. and {M{\'e}nard}, B. and {Zhu}, G.},
    title = "{The Properties of the Cool Circumgalactic Gas Probed
                  with the SDSS, WISE, and GALEX Surveys}",
  journal = {\apj},
archivePrefix = "arXiv",
   eprint = {1404.5301},
     year = 2014,
    month = nov,
   volume = 795,
      eid = {31},
    pages = {31},
      doi = {10.1088/0004-637X/795/1/31},
   adsurl = {http://ui.adsabs.harvard.edu/abs/2014ApJ...795...31L}
}

@ARTICLE{liang18,
       author = {{Liang}, Cameron J. and {Kravtsov}, Andrey V. and {Agertz}, Oscar},
        title = "{Observing the circumgalactic medium of simulated galaxies through synthetic absorption spectra}",
      journal = {\mnras},
         year = 2018,
        month = sep,
       volume = {479},
       number = {2},
        pages = {1822-1835},
          doi = {10.1093/mnras/sty1668},
archivePrefix = {arXiv},
       eprint = {1710.00411},
 primaryClass = {astro-ph.GA},
       adsurl = {https://ui.adsabs.harvard.edu/abs/2018MNRAS.479.1822L}
}

@ARTICLE{lopez20,
       author = {{Lopez}, S. and {Tejos}, N. and {Barrientos}, L.~F. and {Ledoux}, C. and {Sharon}, K. and {Katsianis}, A. and {Florian}, M.~K. and {Rivera-Thorsen}, E. and {Bayliss}, M.~B. and {Dahle}, H. and {Fernandez-Figueroa}, A. and {Gladders}, M.~D. and {Gronke}, M. and {Hamel}, M. and {Pessa}, I. and {Rigby}, J.~R.},
        title = "{Slicing the cool circumgalactic medium along the major axis of a star-forming galaxy at z = 0.7}",
      journal = {\mnras},
         year = 2020,
        month = jan,
       volume = {491},
       number = {3},
        pages = {4442-4461},
          doi = {10.1093/mnras/stz3183},
archivePrefix = {arXiv},
       eprint = {1911.04809},
 primaryClass = {astro-ph.GA},
       adsurl = {https://ui.adsabs.harvard.edu/abs/2020MNRAS.491.4442L}
}

@ARTICLE{lundgren21,
       author = {{Lundgren}, Britt F. and {Creech}, Samantha and {Brammer}, Gabriel and {Kirse}, Nathan and {Peek}, Matthew and {Wake}, David and {York}, Donald G. and {Chisholm}, John and {Erb}, Dawn K. and {Kulkarni}, Varsha P. and {Straka}, Lorrie and {Tremonti}, Christy and {van Dokkum}, Pieter},
        title = "{The Geometry of Cold, Metal-enriched Gas around Galaxies at z {\ensuremath{\sim}} 1.2}",
      journal = {\apj},
         year = 2021,
        month = may,
       volume = {913},
       number = {1},
          eid = {50},
        pages = {50},
          doi = {10.3847/1538-4357/abef6a},
archivePrefix = {arXiv},
       eprint = {2102.10117},
 primaryClass = {astro-ph.GA},
       adsurl = {https://ui.adsabs.harvard.edu/abs/2021ApJ...913...50L}
}

@ARTICLE{marinacci19,
       author = {{Marinacci}, Federico and {Sales}, Laura V. and {Vogelsberger}, Mark and {Torrey}, Paul and {Springel}, Volker},
        title = "{Simulating the interstellar medium and stellar feedback on a moving mesh: implementation and isolated galaxies}",
      journal = {\mnras},
         year = 2019,
        month = nov,
       volume = {489},
       number = {3},
        pages = {4233-4260},
          doi = {10.1093/mnras/stz2391},
archivePrefix = {arXiv},
       eprint = {1905.08806},
 primaryClass = {astro-ph.GA},
       adsurl = {https://ui.adsabs.harvard.edu/abs/2019MNRAS.489.4233M}
}

@ARTICLE{marra21,
       author = {{Marra}, Rachel and {Churchill}, Christopher W. and {Doughty}, Caitlin and {Kacprzak}, Glenn G. and {Charlton}, Jane and {Sameer} and {Nielsen}, Nikole M. and {Ceverino}, Daniel and {Trujillo-Gomez}, Sebastian},
        title = "{Using cosmological simulations and synthetic absorption spectra to assess the accuracy of observationally derived CGM metallicities}",
      journal = {\mnras},
         year = 2021,
        month = dec,
       volume = {508},
       number = {4},
        pages = {4938-4951},
          doi = {10.1093/mnras/stab2896},
archivePrefix = {arXiv},
       eprint = {2108.03254},
 primaryClass = {astro-ph.GA},
       adsurl = {https://ui.adsabs.harvard.edu/abs/2021MNRAS.508.4938M}
}

@ARTICLE{martin19,
       author = {{Martin}, Crystal L. and {Ho}, Stephanie H. and
                  {Kacprzak}, Glenn G. and {Churchill}, Christopher
                  W.},
        title = "{Kinematics of Circumgalactic Gas: Feeding Galaxies
                  and Feedback}",
      journal = {\apj},
         year = "2019",
        month = "Jun",
       volume = {878},
       number = {2},
          eid = {84},
        pages = {84},
          doi = {10.3847/1538-4357/ab18ac},
archivePrefix = {arXiv},
       eprint = {1901.09123},
 primaryClass = {astro-ph.GA},
       adsurl = {https://ui.adsabs.harvard.edu/abs/2019ApJ...878...84M}
}

@ARTICLE{messere26,
       author = {{Messere}, Michael and {Tchernyshyov}, Kirill and {Putman}, Mary E. and {Bryan}, Greg L. and {Werk}, Jessica K. and {Zheng}, Yong and {Schiminovich}, David},
        title = "{Mainly on the Plane: Observing the Extended, Ionized Disks of Milky Way Analogs in IllustrisTNG}",
      journal = {arXiv e-prints},
         year = 2026,
        month = mar,
          eid = {arXiv:2603.22257},
        pages = {arXiv:2603.22257},
          doi = {10.48550/arXiv.2603.22257},
archivePrefix = {arXiv},
       eprint = {2603.22257},
 primaryClass = {astro-ph.GA},
       adsurl = {https://ui.adsabs.harvard.edu/abs/2026arXiv260322257M}
}

@ARTICLE{muratov17,
       author = {{Muratov}, Alexander L. and {Kere{\v{s}}}, Du{\v{s}}an and {Faucher-Gigu{\`e}re}, Claude-Andr{\'e} and {Hopkins}, Philip F. and {Ma}, Xiangcheng and {Angl{\'e}s-Alc{\'a}zar}, Daniel and {Chan}, T.~K. and {Torrey}, Paul and {Hafen}, Zachary H. and {Quataert}, Eliot and {Murray}, Norman},
        title = "{Metal flows of the circumgalactic medium, and the metal budget in galactic haloes}",
      journal = {\mnras},
         year = 2017,
        month = jul,
       volume = {468},
       number = {4},
        pages = {4170-4188},
          doi = {10.1093/mnras/stx667},
archivePrefix = {arXiv},
       eprint = {1606.09252},
 primaryClass = {astro-ph.GA},
       adsurl = {https://ui.adsabs.harvard.edu/abs/2017MNRAS.468.4170M}
}

@ARTICLE{nateghi24GFI,
       author = {{Nateghi}, Hasti and {Kacprzak}, Glenn G. and {Nielsen}, Nikole M. and {Murphy}, Michael T. and {Churchill}, Christopher W. and {Muzahid}, Sowgat and {Sameer} and {Charlton}, Jane C.},
        title = "{Signatures of gas flows - I. Connecting the kinematics of the H I circumgalactic medium to galaxy rotation}",
      journal = {\mnras},
         year = 2024,
        month = sep,
       volume = {533},
       number = {2},
        pages = {1321-1340},
          doi = {10.1093/mnras/stae1843},
archivePrefix = {arXiv},
       eprint = {2311.05164},
 primaryClass = {astro-ph.GA},
       adsurl = {https://ui.adsabs.harvard.edu/abs/2024MNRAS.533.1321N}
}

@ARTICLE{nateghi24GFII,
       author = {{Nateghi}, Hasti and {Kacprzak}, Glenn G. and {Nielsen}, Nikole M. and {Sameer} and {Murphy}, Michael T. and {Churchill}, Christopher W. and {Charlton}, Jane C.},
        title = "{Signatures of gas flows - II. Connecting the kinematics of the multiphase circumgalactic medium to galaxy rotation}",
      journal = {\mnras},
         year = 2024,
        month = oct,
       volume = {534},
       number = {1},
        pages = {930-947},
          doi = {10.1093/mnras/stae2129},
archivePrefix = {arXiv},
       eprint = {2311.05165},
 primaryClass = {astro-ph.GA},
       adsurl = {https://ui.adsabs.harvard.edu/abs/2024MNRAS.534..930N}
}

@ARTICLE{nelson19,
       author = {{Nelson}, Dylan and {Pillepich}, Annalisa and
                  {Springel}, Volker and {Pakmor}, R{\"u}diger and
                  {Weinberger}, Rainer and {Genel}, Shy and {Torrey},
                  Paul and {Vogelsberger}, Mark and {Marinacci},
                  Federico and {Hernquist}, Lars},
        title = "{First results from the TNG50 simulation: galactic
                  outflows driven by supernovae and black hole
                  feedback}",
      journal = {\mnras},
         year = "2019",
        month = "Dec",
       volume = {490},
       number = {3},
        pages = {3234-3261},
          doi = {10.1093/mnras/stz2306},
archivePrefix = {arXiv},
       eprint = {1902.05554},
 primaryClass = {astro-ph.GA},
       adsurl = {https://ui.adsabs.harvard.edu/abs/2019MNRAS.490.3234N}
}

@ARTICLE{magiicat5,
   author = {{Nielsen}, N.~M. and {Churchill}, C.~W. and {Kacprzak},
                  G.~G. and {Murphy}, M.~T. and {Evans}, J.~L.},
    title = "{MAGIICAT V. Orientation of Outflows and Accretion
                  Determine the Kinematics and Column Densities of the
                  Circumgalactic Medium}",
  journal = {\apj},
archivePrefix = "arXiv",
   eprint = {1505.07167},
     year = 2015,
    month = oct,
   volume = 812,
      eid = {83},
    pages = {83},
      doi = {10.1088/0004-637X/812/1/83},
   adsurl = {http://ui.adsabs.harvard.edu/abs/2015ApJ...812...83N},
     note = {({\magiicat} V)}
}

@ARTICLE{peeples19,
       author = {{Peeples}, Molly S. and {Corlies}, Lauren and
                  {Tumlinson}, Jason and {O'Shea}, Brian W. and
                  {Lehner}, Nicolas and {O'Meara}, John M. and {Howk},
                  J. Christopher and {Earl}, Nicholas and {Smith},
                  Britton D. and {Wise}, John H. and {Hummels},
                  Cameron B.},
        title = "{Figuring Out Gas \&amp; Galaxies in Enzo
                  (FOGGIE). I. Resolving Simulated Circumgalactic
                  Absorption at 2 {\ensuremath{\leq}} z
                  {\ensuremath{\leq}} 2.5}",
      journal = {\apj},
         year = "2019",
        month = "Mar",
       volume = {873},
       number = {2},
          eid = {129},
        pages = {129},
          doi = {10.3847/1538-4357/ab0654},
archivePrefix = {arXiv},
       eprint = {1810.06566},
 primaryClass = {astro-ph.GA},
       adsurl = {https://ui.adsabs.harvard.edu/abs/2019ApJ...873..129P}
}

@ARTICLE{peroux20,
       author = {{P{\'e}roux}, C{\'e}line and {Nelson}, Dylan and {van de Voort}, Freeke and {Pillepich}, Annalisa and {Marinacci}, Federico and {Vogelsberger}, Mark and {Hernquist}, Lars},
        title = "{Predictions for the angular dependence of gas mass flow rate and metallicity in the circumgalactic medium}",
      journal = {\mnras},
         year = 2020,
        month = dec,
       volume = {499},
       number = {2},
        pages = {2462-2473},
          doi = {10.1093/mnras/staa2888},
archivePrefix = {arXiv},
       eprint = {2009.07809},
 primaryClass = {astro-ph.GA},
       adsurl = {https://ui.adsabs.harvard.edu/abs/2020MNRAS.499.2462P}
}

@ARTICLE{peroux16,
       author = {{P{\'e}roux}, C{\'e}line and {Quiret}, Samuel and
                  {Rahmani}, Hadi and {Kulkarni}, Varsha P. and
                  {Epinat}, Benoit and {Milliard}, Bruno and {Straka},
                  Lorrie A. and {York}, Donald G. and {Rahmati},
                  Alireza and {Contini}, Thierry},
        title = "{A SINFONI integral field spectroscopy survey for
                  galaxy counterparts to damped Lyman
                  {\ensuremath{\alpha}} systems - VI. Metallicity and
                  geometry as gas flow probes}",
      journal = {\mnras},
         year = "2016",
        month = "Mar",
       volume = {457},
       number = {1},
        pages = {903-916},
          doi = {10.1093/mnras/stw016},
archivePrefix = {arXiv},
       eprint = {1601.02796},
 primaryClass = {astro-ph.GA},
       adsurl = {https://ui.adsabs.harvard.edu/abs/2016MNRAS.457..903P}
}

@ARTICLE{pointon19,
       author = {{Pointon}, Stephanie K. and {Kacprzak}, Glenn G. and
                  {Nielsen}, Nikole M. and {Muzahid}, Sowgat and
                  {Murphy}, Michael T. and {Churchill}, Christopher
                  W. and {Charlton}, Jane C.},
        title = "{Relationship between the Metallicity of the
                  Circumgalactic Medium and Galaxy Orientation}",
      journal = {\apj},
         year = "2019",
        month = "Sep",
       volume = {883},
       number = {1},
          eid = {78},
        pages = {78},
          doi = {10.3847/1538-4357/ab3b0e},
archivePrefix = {arXiv},
       eprint = {1907.05557},
 primaryClass = {astro-ph.GA},
       adsurl = {https://ui.adsabs.harvard.edu/abs/2019ApJ...883...78P}
}

@ARTICLE{prochaska17,
       author = {{Prochaska}, J. Xavier and {Werk}, Jessica K. and {Worseck}, G{\'a}bor and {Tripp}, Todd M. and {Tumlinson}, Jason and {Burchett}, Joseph N. and {Fox}, Andrew J. and {Fumagalli}, Michele and {Lehner}, Nicolas and {Peeples}, Molly S. and {Tejos}, Nicolas},
        title = "{The COS-Halos Survey: Metallicities in the Low-redshift Circumgalactic Medium}",
      journal = {\apj},
         year = 2017,
        month = mar,
       volume = {837},
       number = {2},
          eid = {169},
        pages = {169},
          doi = {10.3847/1538-4357/aa6007},
archivePrefix = {arXiv},
       eprint = {1702.02618},
 primaryClass = {astro-ph.GA},
       adsurl = {https://ui.adsabs.harvard.edu/abs/2017ApJ...837..169P}
}

@ARTICLE{rahmani18,
       author = {{Rahmani}, Hadi and {P{\'e}roux}, C{\'e}line and
                  {Augustin}, Ramona and {Husemann}, Bernd and
                  {Kacprzak}, Glenn G. and {Kulkarni}, Varsha and
                  {Milliard}, Bruno and {M{\o}ller}, Palle and
                  {Pettini}, Max and {Straka}, Lorrie and {Vernet},
                  Jo{\"e}l and {York}, Donald G.},
        title = "{Observational signatures of a warped disk associated
                  with cold-flow accretion}",
      journal = {\mnras},
         year = 2018,
        month = Feb,
       volume = {474},
        pages = {254-270},
          doi = {10.1093/mnras/stx2726},
       adsurl = {https://ui.adsabs.harvard.edu/#abs/2018MNRAS.474..254R}
}

@ARTICLE{sameer21,
       author = {{Sameer} and {Charlton}, Jane C. and {Norris}, Jackson M. and {Gebhardt}, Matthew and {Churchill}, Christopher W. and {Kacprzak}, Glenn G. and {Muzahid}, Sowgat and {Narayanan}, Anand and {Nielsen}, Nikole M. and {Richter}, Philipp and {Wakker}, Bart P.},
        title = "{Cloud-by-cloud, multiphase, Bayesian modelling: application to four weak, low-ionization absorbers}",
      journal = {\mnras},
         year = 2021,
        month = feb,
       volume = {501},
       number = {2},
        pages = {2112-2139},
          doi = {10.1093/mnras/staa3754},
archivePrefix = {arXiv},
       eprint = {2012.00021},
 primaryClass = {astro-ph.GA},
       adsurl = {https://ui.adsabs.harvard.edu/abs/2021MNRAS.501.2112S}
}

@ARTICLE{sameer22,
       author = {{Sameer} and {Charlton}, Jane C. and {Kacprzak}, Glenn G. and {Narayanan}, Anand and {Sankar}, Sriram and {Richter}, Philipp and {Wakker}, Bart P. and {Nielsen}, Nikole M. and {Churchill}, Christopher W.},
        title = "{Probing the physicochemical properties of the Leo Ring and the Leo I group}",
      journal = {\mnras},
         year = 2022,
        month = mar,
       volume = {510},
       number = {4},
        pages = {5796-5820},
          doi = {10.1093/mnras/stac052},
archivePrefix = {arXiv},
       eprint = {2201.02631},
 primaryClass = {astro-ph.GA},
       adsurl = {https://ui.adsabs.harvard.edu/abs/2022MNRAS.510.5796S}
}

@ARTICLE{sameer24,
       author = {{Sameer} and {Charlton}, Jane C. and {Wakker}, Bart P. and {Kacprzak}, Glenn G. and {Nielsen}, Nikole M. and {Churchill}, Christopher W. and {Richter}, Philipp and {Muzahid}, Sowgat and {Ho}, Stephanie H. and {Nateghi}, Hasti and {Rosenwasser}, Benjamin and {Narayanan}, Anand and {Ganguly}, Rajib},
        title = "{Cloud-by-cloud multiphase investigation of the circumgalactic medium of low-redshift galaxies}",
      journal = {\mnras},
         year = 2024,
        month = jun,
       volume = {530},
       number = {4},
        pages = {3827-3854},
          doi = {10.1093/mnras/stae962},
archivePrefix = {arXiv},
       eprint = {2403.05617},
 primaryClass = {astro-ph.GA},
       adsurl = {https://ui.adsabs.harvard.edu/abs/2024MNRAS.530.3827S}
}

@ARTICLE{schroetter16,
   author = {{Schroetter}, I. and {Bouch{\'e}}, N. and {Wendt}, M. and
                  {Contini}, T. and {Finley}, H. and {Pell{\'o}},
                  R. and {Bacon}, R. and {Cantalupo}, S. and {Marino},
                  R.~A. and {Richard}, J. and {Lilly}, S.~J. and
                  {Schaye}, J. and {Soto}, K. and {Steinmetz}, M. and
                  {Straka}, L.~A. and {Wisotzki}, L.  },
    title = "{Muse Gas Flow and Wind (MEGAFLOW). I. First MUSE Results
                  on Background Quasars}",
  journal = {\apj},
archivePrefix = "arXiv",
   eprint = {1605.03412},
     year = 2016,
    month = dec,
   volume = 833,
      eid = {39},
    pages = {39},
      doi = {10.3847/1538-4357/833/1/39},
   adsurl = {http://ui.adsabs.harvard.edu/abs/2016ApJ...833...39S}
}

@ARTICLE{schroetter19,
       author = {{Schroetter}, Ilane and {Bouch{\'e}}, Nicolas F. and
                  {Zabl}, Johannes and {Contini}, Thierry and {Wendt},
                  Martin and {Schaye}, Joop and {Mitchell}, Peter and
                  {Muzahid}, Sowgat and {Marino}, Raffaella A. and
                  {Bacon}, Roland and {Lilly}, Simon J. and {Richard},
                  Johan and {Wisotzki}, Lutz},
        title = "{MusE GAs FLOw and Wind (MEGAFLOW) - III. Galactic
                  wind properties using background quasars}",
      journal = {\mnras},
         year = "2019",
        month = "Dec",
       volume = {490},
       number = {3},
        pages = {4368-4381},
          doi = {10.1093/mnras/stz2822},
archivePrefix = {arXiv},
       eprint = {1907.09967},
 primaryClass = {astro-ph.GA},
       adsurl = {https://ui.adsabs.harvard.edu/abs/2019MNRAS.490.4368S}
}

@ARTICLE{steidel02,
   author = {{Steidel}, C.~C. and {Kollmeier}, J.~A. and {Shapley},
                  A.~E. and {Churchill}, C.~W. and {Dickinson}, M. and
                  {Pettini}, M.},
    title = "{The Kinematic Connection between absorbing Gas toward
                  QSOs and Galaxies at Intermediate Redshift}",
  journal = {\apj},
   eprint = {astro-ph/0201353},
     year = 2002,
    month = may,
   volume = 570,
    pages = {526-542},
      doi = {10.1086/339792},
   adsurl = {http://ui.adsabs.harvard.edu/abs/2002ApJ...570..526S}
}

@ARTICLE{stern24,
       author = {{Stern}, Jonathan and {Fielding}, Drummond and {Hafen}, Zachary and {Su}, Kung-Yi and {Naor}, Nadav and {Faucher-Gigu{\`e}re}, Claude-Andr{\'e} and {Quataert}, Eliot and {Bullock}, James},
        title = "{Accretion onto disc galaxies via hot and rotating CGM inflows}",
      journal = {\mnras},
         year = 2024,
        month = may,
       volume = {530},
       number = {2},
        pages = {1711-1731},
          doi = {10.1093/mnras/stae824},
archivePrefix = {arXiv},
       eprint = {2306.00092},
 primaryClass = {astro-ph.GA},
       adsurl = {https://ui.adsabs.harvard.edu/abs/2024MNRAS.530.1711S}
}

@ARTICLE{stewart11,
   author = {{Stewart}, K.~R. and {Kaufmann}, T. and {Bullock},
                  J.~S. and {Barton}, E.~J. and {Maller}, A.~H. and
                  {Diemand}, J. and {Wadsley}, J.  },
    title = "{Orbiting Circumgalactic Gas as a Signature of
                  Cosmological Accretion}",
  journal = {\apj},
archivePrefix = "arXiv",
   eprint = {1103.4388},
 primaryClass = "astro-ph.CO",
     year = 2011,
    month = sep,
   volume = 738,
      eid = {39},
    pages = {39},
      doi = {10.1088/0004-637X/738/1/39},
   adsurl = {http://ui.adsabs.harvard.edu/abs/2011ApJ...738...39S}
}

\bsp	
\label{lastpage}
\end{document}